\documentclass[aps,prx,twocolumn,superscriptaddress,showpacs,showkeys,10pt,longbibliography]{revtex4-1}
\usepackage{graphicx}
\usepackage{float}
\usepackage{textcomp}
\usepackage{color}
\usepackage{dcolumn}
\usepackage{multirow}
\usepackage{physics}
\usepackage{amsmath}
\usepackage{amssymb}
\usepackage{siunitx}
\newcommand{\refeq}[1]{Eq.~(\ref{#1})}
\newcommand{\reffig}[1]{Fig.~\ref{#1}}

\newcommand{\be}{\begin{equation}}
\newcommand{\ee}{\end{equation}}
\newcommand{\bea}{\begin{eqnarray}}
\newcommand{\eea}{\end{eqnarray}}
\usepackage[colorlinks]{hyperref}
\hypersetup{
	plainpages=true,
	breaklinks=true,
	hypertexnames=false,
	pageanchor=true,
	colorlinks=true,
	linkcolor={blue},
	citecolor={blue},
	urlcolor={blue},
	anchorcolor={black}
}

\begin{document}

\title{A universal quantum gate set for transmon qubits with strong ZZ interactions}
\author{Junling Long}
\email{Junling.Long@Colorado.edu}
\affiliation{Department of Physics, University of Colorado, Boulder, Colorado 80309, USA}
\author{Tongyu Zhao}
\email{Tongyu.Zhao@Colorado.edu}
\affiliation{Department of Physics, University of Colorado, Boulder, Colorado 80309, USA}
\author{Mustafa Bal}
\affiliation{Department of Physics, University of Colorado, Boulder, Colorado 80309, USA}
\author{Ruichen Zhao}
\affiliation{Department of Physics, University of Colorado, Boulder, Colorado 80309, USA}
\author{George S. Barron}
\affiliation{Department of Physics, Virginia Tech, Blacksburg, VA 24061, USA}
\author{Hsiang-sheng Ku}
\affiliation{National Institute of Standards and Technology, Boulder, Colorado 80305, USA}
\author{Joel A. Howard}
\affiliation{Department of Physics, Colorado School of Mines, Golden, Colorado 80401, USA}
\author{Xian Wu}
\affiliation{Department of Physics, University of Colorado, Boulder, Colorado 80309, USA}
\author{Corey Rae H. McRae}
\affiliation{Department of Physics, University of Colorado, Boulder, Colorado 80309, USA}
\author{Xiu-Hao Deng}
\affiliation{Shenzhen Institute of Quantum Science and Engineering, Southern University of Science and Technology, Shenzhen, Guangdong 518055, China}
\author{Guilhem J. Ribeill}
\affiliation{Quantum Engineering and Computing, Raytheon BBN Technologies, Cambridge, MA 02138, USA.}
\author{Meenakshi Singh}
\affiliation{Department of Physics, Colorado School of Mines, Golden, Colorado 80401, USA}
\author{Thomas A. Ohki}
\affiliation{Quantum Engineering and Computing, Raytheon BBN Technologies, Cambridge, MA 02138, USA.}
\author{Edwin Barnes}
\affiliation{Department of Physics, Virginia Tech, Blacksburg, VA 24061, USA}
\author{Sophia E. Economou}
\affiliation{Department of Physics, Virginia Tech, Blacksburg, VA 24061, USA}
\author{David P. Pappas}
\affiliation{National Institute of Standards and Technology, Boulder, Colorado 80305, USA}

\date{\today}

\begin{abstract}
High-fidelity single- and two-qubit gates are essential building blocks for a fault-tolerant quantum computer. While there has been much progress in suppressing single-qubit gate errors in superconducting qubit systems, two-qubit gates still suffer from error rates that are orders of magnitude higher. One limiting factor is the residual ZZ-interaction, which originates from a coupling between computational states and higher-energy states. While this interaction is usually viewed as a nuisance, here we experimentally demonstrate that it can be exploited to produce a universal set of fast single- and two-qubit entangling gates in a coupled transmon qubit system. To implement arbitrary single-qubit rotations, we design a new protocol called the two-axis gate that is based on a three-part composite pulse. It rotates a single qubit independently of the state of the other qubit despite the strong ZZ-coupling. We achieve single-qubit gate fidelities as high as 99.1\% (for $\widetilde{\ket{01}}\leftrightarrow\widetilde{\ket{11}}$ transition) from randomized benchmarking measurements. We then demonstrate both a CZ gate and a CNOT gate. Because the system has a strong ZZ-interaction, a CZ gate can be achieved by letting the system freely evolve for a gate time $t_g=\SI{53.8}{\ns}$. To design the CNOT gate, we utilize an analytical microwave pulse shape based on the SWIPHT protocol for realizing fast, low-leakage gates. We obtain fidelities of 94.6\% and 97.8\% for the CNOT and CZ gates respectively from quantum progress tomography. 
\end{abstract}


\maketitle

\section{Introduction}
Superconducting qubits hold promise as the main building blocks of a future fault-tolerant quantum computer~\cite{gambetta2017building,krantz2019quantum,devoret2013superconducting}. In recent years, advances in both coherence times~\cite{kjaergaard2019superconducting} and quantum gate fidelities~\cite{barends2014superconducting,mckay2019three,hong2020demonstration}, together with the inherent scalability of superconducting circuit architectures, have enabled demonstrations of superconducting quantum computers comprised of tens of qubits~\cite{cross2019validating,arute2019quantum,otterbach2017unsupervised}. 
While single-qubit gate errors are routinely on the order of 0.01\%~\cite{motzoi2009simple,sheldon2016characterizing,kelly2014optimal}, two-qubit gates across different qubit architectures have exhibited higher error rates, ranging from 0.1\% up to the few-percent level~\cite{kjaergaard2019superconducting,foxen2020demonstrating,barends2014superconducting,kjaergaard2020quantum,gambetta2017building,mckay2019three,krinner2020demonstration,xu2020high,collodo2020implementation,mckay2016universal}. Thus, two-qubit gates remain a bottleneck in the performance of existing devices. 

One significant source of two-qubit gate errors is the residual ZZ interaction between coupled qubits~\cite{mckay2019three}. The ZZ interaction, also known as ZZ coupling or crosstalk, originates from a mixing between computational states and higher-energy states of the coupled qubits. It shifts the resonance frequency of one qubit conditionally on the state of another qubit. As a result, spurious phases are accumulated during gates, reducing fidelities. Some efforts have been devoted to suppressing the ZZ coupling using tunable couplers~\cite{mundada2019suppression,Li2020_tunable_coupler} or hybrid superconducting qubits~\cite{ku2020suppression}.

With the recent progress in tunable couplers \cite{mundada2019suppression,Li2020_tunable_coupler,collodo2020implementation}, new opportunities open up for scalable architectures. In this context, it has been realized and demonstrated that the ZZ interaction can in fact be exploited for the implementation of a maximally entangling CZ gate \cite{collodo2020implementation}. This approach is very promising for scalability, as the qubits can be brought into the strong ZZ coupling regime to generate fast gates, while in the idling regime crosstalk is suppressed. Since tuning can be costly in terms of time and possibly coherence, it would be beneficial for the performance of the device to maximize the use of this strongly coupled regime by developing additional quantum gates that can be implemented within it. For example, a CNOT gate, despite being locally equivalent to a CZ, can be directly implemented using microwave fields \cite{economou2015analytical,deng2017robustness}, eliminating the need for additional single-qubit rotations. The development of a broader toolset of gates applicable in the nonzero ZZ coupling regime is also important for cases where this coupling cannot be completely switched off, which is the generic situation in superconducting circuits. 

In this work, we develop and demonstrate a universal set of fast gates that operate in the regime of strong ZZ interactions between capacitively coupled superconducting transmon qubits. To implement arbitrary single-qubit gates in this regime, we introduce a family of three-part composite pulses that effectively freeze the ZZ coupling so that one qubit is rotated independently of the state of another. We call these two-axis gates (TAGs). Randomized benchmarking is performed on each qubit to characterize the TAG fidelity, which we find to be 99.1\%. We also demonstrate two types of maximally entangling gates: CZ and CNOT. The CZ gate is implemented via free evolution under the ZZ interaction, yielding a gate time of 53.8 ns. For the CNOT gate, we employ the SWIPHT protocol \cite{economou2015analytical,deng2017robustness,barron2020microwave}, which amounts to cleverly designing the pulse such that it acts on two transitions (the target and the nearest competing transition) in such a way that the net effect is the implementation of the target gate. This allows for a much faster gate compared to resolving the transitions. We create a maximally entangled state with the CNOT gate and obtain a fidelity of 98.2\%, confirmed using quantum state tomography. Finally, we characterize the CZ and CNOT gates using quantum process tomography. The measured average gate fidelity is about 94.6\% (97.8\%) for the CNOT (CZ) gate. 

This paper is organized as follows. In Sec.~\ref{sec:device}, we provide details about the transmon device and the effective ZZ Hamiltonian. Sec.~\ref{sec:singlequbit} presents our two-axis gate designs for implementing arbitrary single-qubit rotations in the strong ZZ coupling regime. In Sec.~\ref{sec:twoqubit}, we describe how we implement our CZ and CNOT gates, including a review of the SWIPHT protocol we use for the latter. Our randomized benchmarking measurements for the two-axis gates are presented in Sec.~\ref{sec:randomizedbenchmarking}, while our quantum state tomography and process tomography results for our two-qubit entangling gates are shown in Secs.~\ref{sec:qst} and \ref{sec:qpt}, respectively. We conclude in Sec.~\ref{sec:conclusions}. Three appendices contain additional information about the spectrum of the system and relaxation and dephasing times, along with a detailed derivation of the effective ZZ interaction for both capacitively coupled and resonator-coupled transmons.

\section{Device}\label{sec:device}
\begin{figure*}[t]
	\includegraphics[width=14cm]{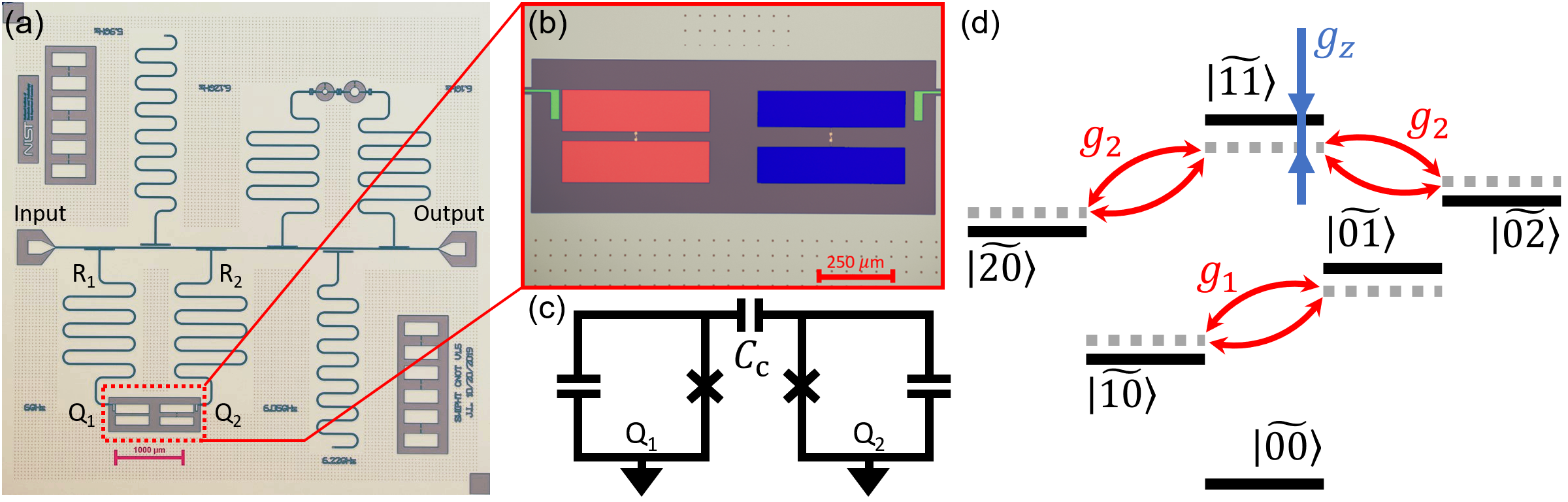}
	\caption[ZZ-coupled transmon qubits]{(a) Optical micrograph of the full chip including qubits, readout resonators, test Josephson junctions, and test resonators. (b) Zoomed-in view of the two floating qubits (DUT) used in this experiment. Each qubit consists of two identical pads (red for the left qubit and blue for the right qubit) and a Josephson junction connecting the two pads. Each qubit is coupled to its own readout resonator (green). (c) Grounded circuit model of the capacitively coupled DUT. See Ref.~\cite{Junling2020Superconducting} for how to obtain the equivalent grounded circuit model of floating qubits. (d) Energy level diagram of the two-qubit system. Gray dashed lines are uncoupled two-qubit energy levels. Solid lines represent the diagonalized energy levels of the two-qubit system with the coupling on. The red bidirectional curved arrows denote the swapping couplings between excitation-preserved states, and the coupling strengths are given by $g_{1}$ and $g_{2}$. $g_{z}$ denotes the shift of $\ket{11}$ due to the couplings.}
\label{ZZCoupledTransmon}
\end{figure*}

The sample chip is shown in \reffig{ZZCoupledTransmon} (a) and (b), where optical images of the full device and a zoomed-in view of the two qubits (red and blue) are shown, respectively. The device-under-test (DUT) consists of two floating transmon qubits \cite{Koch2007} capacitively coupled to each other. Each transmon qubit is coupled to a coplanar waveguide readout resonator. The two readout resonators are coupled to the feedline in a hanger-style. Readout and control signals are all sent through the feedline. The capacitive components of each transmon are comprised of two identical rectangular pads [red (blue) for the first (second) qubit, as shown in \reffig{ZZCoupledTransmon}(b)]. Each pair of pads is connected by an Al/AlO$_x$/Al Josephson junction that is fabricated with an overlap technique \cite{wu2017overlap}. The full chip (except for the Josephson junctions) is made of 100~nm thick Niobium superconducting films. In this particular DUT, the two transmon qubits have frequencies of $\omega_{1}/2\pi=\SI{5.075}{\GHz}$ and $\omega_{2}/2\pi=\SI{5.310}{\GHz}$, and anharmonicities of $\alpha_{1}/2\pi=\SI{-260}{\MHz}$ and $\alpha_{2}/2\pi=\SI{-340}{\MHz}$ (Appendix \ref{app:spectrum}). The relaxation and coherence times of the two qubits are measured to be $T_{1}^{(1)}=\SI{76.98}{\us}$, $T_{2}^{*(1)}=\SI{50.65}{\us}$, $T_{1}^{(2)}=\SI{79.71}{\us}$, and $T_{2}^{*(2)}=\SI{17.09}{\us}$ (see Appendix \ref{app:decoherence}). The equivalent grounded circuit model of the two capacitively coupled transmon qubits is shown in \reffig{ZZCoupledTransmon}(c). The Gaussian elimination method described in Ref.~\cite{Junling2020Superconducting} was applied to convert the full floating circuit model to this simplified grounded circuit model. Note that the circuit of the readout resonators is not shown.

The level diagram of the DUT is shown in \reffig{ZZCoupledTransmon}(d). Two-qubit states are labeled as $\ket{mn}$, where $m$ and $n$ in the ket represent the $m$th state of the first qubit and $n$th state of the second qubit, respectively. The gray dashed lines in \reffig{ZZCoupledTransmon}(d) are the uncoupled two-qubit states, while the solid lines are the dressed two-qubit states with the always-on capacitive coupling. We only retain states involving up to two total excitations. The coupling between the $\ket{10}$ and $\ket{01}$ states is given by $g_{1}$. This coupling causes the two states to repel each other. As a result, the dressed qubit frequency shifts down (up) compared to the uncoupled qubit frequency of the first (second) qubit.

The couplings in the two-excitation manifold are between the $\ket{20}\leftrightarrow\ket{11}$, and $\ket{11}\leftrightarrow\ket{02}$ states, with the same strength, $g_{2}\approx \sqrt{2}g_{1}$. As a result of the two couplings, the dressed energy level $\widetilde{\ket{11}}$ is shifted up from the uncoupled $\ket{11}$ state. This makes the frequencies of the transitions $\widetilde{\ket{01}}\leftrightarrow\widetilde{\ket{11}}$ and $\widetilde{\ket{10}}\leftrightarrow\widetilde{\ket{11}}$ higher than those of $\widetilde{\ket{00}}\leftrightarrow\widetilde{\ket{10}}$ and $\widetilde{\ket{00}}\leftrightarrow\widetilde{\ket{01}}$, respectively, i.e., this is a state-dependent shift in qubit frequencies. In the computational subspace, defined by the dressed qubit states $\widetilde{\ket{00}}$, $\widetilde{\ket{01}}$, $\widetilde{\ket{10}}$, and $\widetilde{\ket{11}}$, the two-qubit system is described by the following Hamiltonian (see Appendix \ref{app:ZZcoupling}),
\begin{equation}
\begin{split}
H_{q}/\hbar = (\omega_{1}+\frac{g_{z}}{2})\frac{\sigma_{z}^{(1)}}{2} + (\omega_{2}+\frac{g_{z}}{2})\frac{\sigma_{z}^{(2)}}{2}+g_{z}\frac{\sigma_{z}^{(1)}}{2}\frac{\sigma_{z}^{(2)}}{2},
\end{split}
\label{2QZZHamiltonian}
\end{equation}
where $\sigma_{z}^{(i)}=(\widetilde{\ket{1}}\widetilde{\bra{1}}-\widetilde{\ket{0}}\widetilde{\bra{0}})$ is the atomic inversion operator for the $i^{th}$ qubit; $\omega_{1}$ ($\omega_{2}$) is the frequency of the first (second) qubit while the other qubit is in its ground state, i.e., it is the transition frequency of $\widetilde{\ket{00}}\leftrightarrow\widetilde{\ket{10}}$ ( $\widetilde{\ket{00}}\leftrightarrow\widetilde{\ket{01}}$); $g_{z}$ is the effective coupling strength determined by the amount of the up-shift of state $\widetilde{\ket{11}}$. We obtain a value of $g_{z}/2\pi=9.29$~MHz from spectroscopy measurements (see Appendix \ref{app:spectrum}). As shown in \refeq{2QZZHamiltonian}, the coupling between the two qubits in the computational subspace (dressed basis) is ZZ, but it originates from the transverse XX coupling between level $\ket{11}$ and the higher levels $\ket{20}$ and $\ket{02}$. As shown in Appendix~\ref{app:ZZcoupling}, Eq.~\ref{2QZZHamiltonian} can originate from either a direct, capacitive coupling or a resonator-mediated coupling between two transmons.

\section{Arbitrary single-qubit rotations}\label{sec:singlequbit}
Performing arbitrary single-qubit rotations is not a trivial task in a two-qubit system with strong ZZ coupling because the frequency of one qubit is conditional on the state of the other qubit. Here, we design a new type of single-qubit gate, called the two-axis gate (TAG), that unconditionally drives arbitrary single-qubit rotations in this ZZ-coupled, two-qubit system.

\begin{figure}[!htb]
    \begin{center}
	\includegraphics[width=8cm]{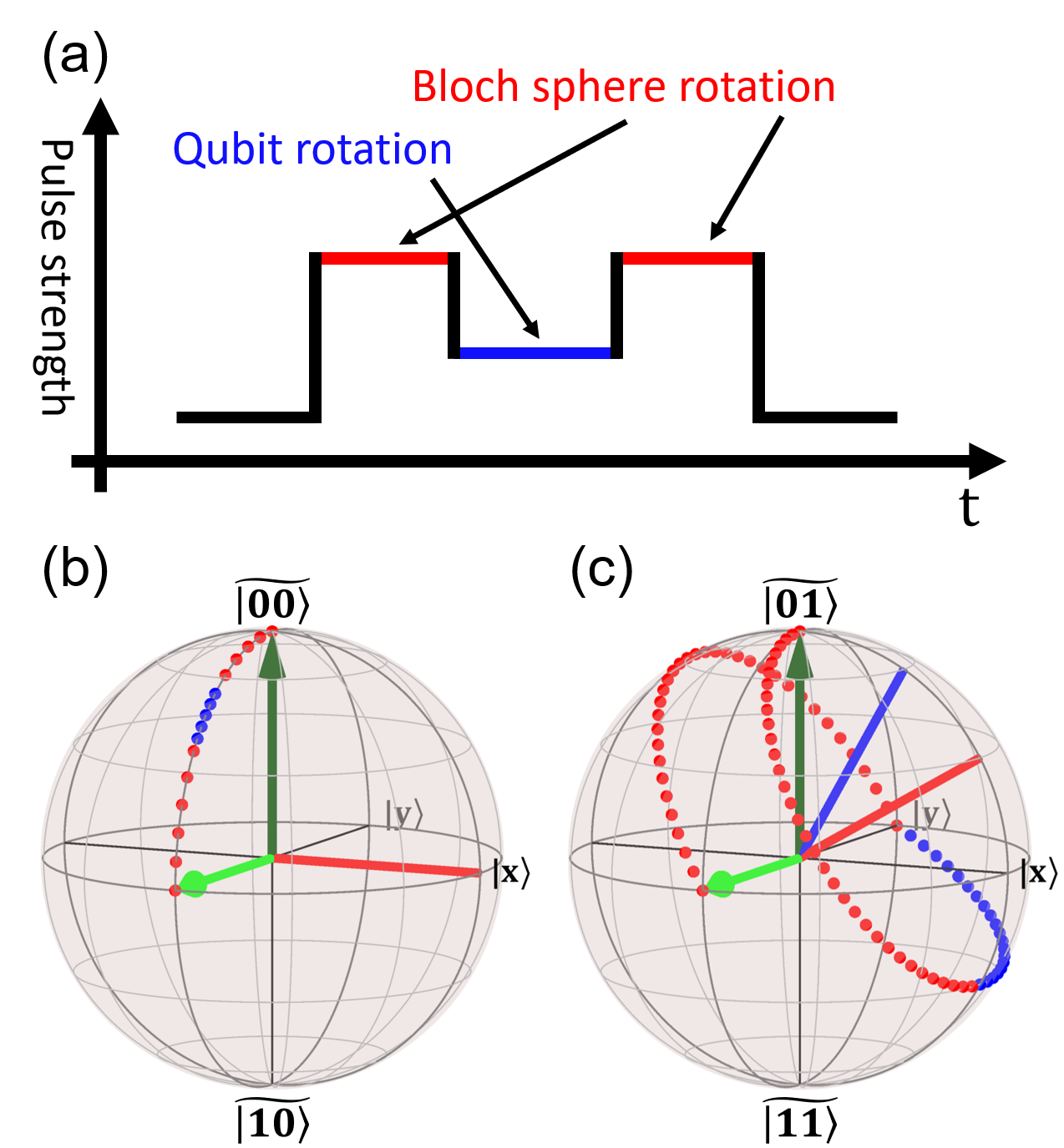}
    \end{center}
	\caption[Square shaped two-axis gate]{(a) A square shaped two-axis gate that implements a $X(\pi/2)$ on the first qubit. (b) and (c) show the trajectories of the state vector for the $\widetilde{\ket{00}}\leftrightarrow\widetilde{\ket{10}}$ and $\widetilde{\ket{01}}\leftrightarrow\widetilde{\ket{11}}$ transitions, respectively. In (b) and (c) the initial state is $\widetilde{\ket{00}}$, as denoted by the dark green arrow, while the final state is $\ket{-y}=(\widetilde{\ket{00}}-i\widetilde{\ket{10}})/\sqrt{2}$, which is denoted by the light green arrow. The red (blue) dots in (a) and (c) denote the trajectories corresponding to the BR (QR) pulses.}
\label{TwoAxesGate_trajectory}
\end{figure}

To explain how TAGs work, we take a X($\pi/2$) rotation on the first qubit as an example. 
As shown in \reffig{TwoAxesGate_trajectory}(a), TAGs are implemented with three-part composite pulses. Note that the TAG protocol can be implemented using any pulse shapes for each of the three pieces. Here, we use square pulses in our experiments to facilitate pulse tune up. The first and third pulses (red) are identical, and we refer to them as Bloch sphere rotation (BR) pulses. The second pulse (blue) is called the qubit rotation (QR) pulse. For our X($\pi/2$) rotation example, the center frequency of the TAG drive is set to be on resonance with the $\widetilde{\ket{00}}\leftrightarrow\widetilde{\ket{10}}$ transition. In the frame rotating at this frequency, the rotation axes for the three pulses of the TAG are all along the $x$-axis, as shown in \reffig{TwoAxesGate_trajectory}(b). The pulse strengths and durations are chosen such that all three pulses combine to rotate the state vector around the $x$-axis by $\pi/2$ and drive the original state $\widetilde{\ket{00}}$ to the final state $\ket{-y}=(\widetilde{\ket{00}}-i\widetilde{\ket{10}})/\sqrt{2}$.

Since the center frequency is detuned from the $\widetilde{\ket{01}}\leftrightarrow\widetilde{\ket{11}}$ transition by $g_z$ as described by \refeq{2QZZHamiltonian}, the rotation axes of the BR and QR are both tilted from the $x$-axis in the $xz$ plane by $\phi_{BR}$  and $\phi_{QR}$, respectively, as shown in \reffig{TwoAxesGate_trajectory}(c). The BR pulses can be treated as a rotation of the Bloch sphere around the blue axis in the opposite direction. This is the reason why we call it a Bloch sphere rotation. When the first BR pulse is applied, we require it to rotate the Bloch sphere such that the $x$-axis after the rotation coincides with the QR axis, i.e., the blue axis in \reffig{TwoAxesGate_trajectory}(c). To achieve this, two conditions need to be satisfied. First, the red axis needs to be the bisector of the angle formed between the blue- and $x$-axis, i.e., $\phi_{QR}=2\phi_{BR}$. Second, the rotation angle of the BR pulse needs to be $\pi$ or $-\pi$. In the rotated Bloch sphere, since the QR (blue) axis is on the $x$-axis, we can perform a rotation around the $x$-axis by $\pi/2$. After the QR pulse, we apply the BR pulse again to rotate the Bloch sphere around the BR (red) axis by another $\pi$ angle to get back to the original Bloch sphere. 
Treating a BR pulse as a rotation of the Bloch sphere is a good way to understand the TAG. If we treat all the pulses in the TAG as rotations of a state vector, we can see the trajectories on both Bloch spheres. This is shown in \reffig{TwoAxesGate_trajectory}(c). When the first BR pulse is applied, the rotated state vector becomes perpendicular to the blue axis. Then, the QR pulse is applied to rotate the state vector by $\Theta=\pi/2$. Finally, the BR pulse is applied again to bring the state to $\ket{-y}$.

Similarly, we can design TAGs that rotate the state of one qubit around the $x$-axis by other arbitrary angles. To perform a $y$-rotation with the TAG, one can just add a $90^{\circ}$ global phase shift to the TAG. Arbitrary rotations can be performed by combinations of $x$ and $y$-rotations.

\section{Two-qubit entangling gates}\label{sec:twoqubit}
In this section, we describe how we realize two different two-qubit entangling gates, a CZ gate and a generalized CNOT gate, on our device.

\subsection{A controlled-Z gate by free evolution}
With the ZZ coupling, free evolution of the system is, in fact, a controlled-phase gate. By transforming the Hamiltonian in \refeq{2QZZHamiltonian} to the rotating frame of $\omega_1$ and $\omega_2$ for the first and second qubit, respectively, and adding a constant energy shift $g_z/4$, one obtains
\begin{equation}
    H^{rot}_{q}/\hbar=g_{z}\widetilde{\ket{11}}\widetilde{\bra{11}}.
\end{equation}
Note that we used the fact that $\sigma_{z}^{(i)}$ is defined in the dressed two-qubit basis to obtain this result. Also note that adding a constant energy shift to the Hamiltonian only produces a global phase; it does not change the dynamics of the two qubits. The time evolution operator of this Hamiltonian above is given by
\begin{equation}
\begin{split}
    U(t)&=e^{-iH^{rot}_{q}t/\hbar}=e^{-i g_{z}t\widetilde{\ket{11}}\widetilde{\bra{11}}}
    \\
    &=\begingroup 
\setlength\arraycolsep{8pt}
\renewcommand{\arraystretch}{1}
    \begin{pmatrix}
    1&0&0&0\\
    0&1&0&0\\
    0&0&1&0\\
    0&0&0&e^{-i g_{z}t}
    \end{pmatrix}.
\endgroup
\end{split}
\end{equation}
This is a standard controlled-phase gate with the phase given by $-g_{z}t$. To realize a CZ gate, we can just let the system evolve for time $t=\frac{\pi}{g_z}$, in which case the evolution operator becomes
\begin{equation}
U_{\text{CZ}}=U\left(\frac{\pi}{g_z}\right)=
\begingroup 
\setlength\arraycolsep{8pt}
\renewcommand{\arraystretch}{1}
    \begin{pmatrix}
    1&0&0&0\\
    0&1&0&0\\
    0&0&1&0\\
    0&0&0&-1
    \end{pmatrix}.
\endgroup
\end{equation}
The CZ gate time is given by $\frac{\pi}{g_{z}}=53.8$~ns. We analyze the performance of this gate using quantum process tomography in Sec.~\ref{sec:qpt}. We note that a similar CZ gate based on free evolution in the strong ZZ coupling regime, achieved via a tunable coupler, was recently demonstrated \cite{collodo2020implementation}.

\subsection{A generalized controlled-NOT gate based on the SWIPHT protocol}
\label{A generalized controlled-NOT gate based on the SWIPHT protocol}
In addition to the free-evolution CZ gate, it is also possible to implement a CNOT gate in the strong ZZ coupling regime using a shaped microwave pulse. A protocol known as SWIPHT (speeding up wave forms by inducing phases to harmful transitions), provides a general framework for designing fast, high-fidelity two-qubit entangling gates~\cite{economou2015analytical,deng2017robustness,barron2020microwave}. The SWIPHT protocol is based on reverse-engineering analytical pulse shapes that implement two-qubit entangling gates~\cite{barnes2012analytically}. SWIPHT was first demonstrated experimentally for a two-transmon system in the weak ZZ coupling regime~\cite{Premaratne2019}. The SWIPHT protocol was originally proposed for two transmon qubits coupled with a bus resonator, but it applies to any two-qubit system with ZZ coupling in the dressed qubit basis. Here, we employ SWIPHT to implement a generalized CNOT gate for our two-qubit system. 

\begin{figure}[!htb]
    \begin{center}
	\includegraphics[width=8cm]{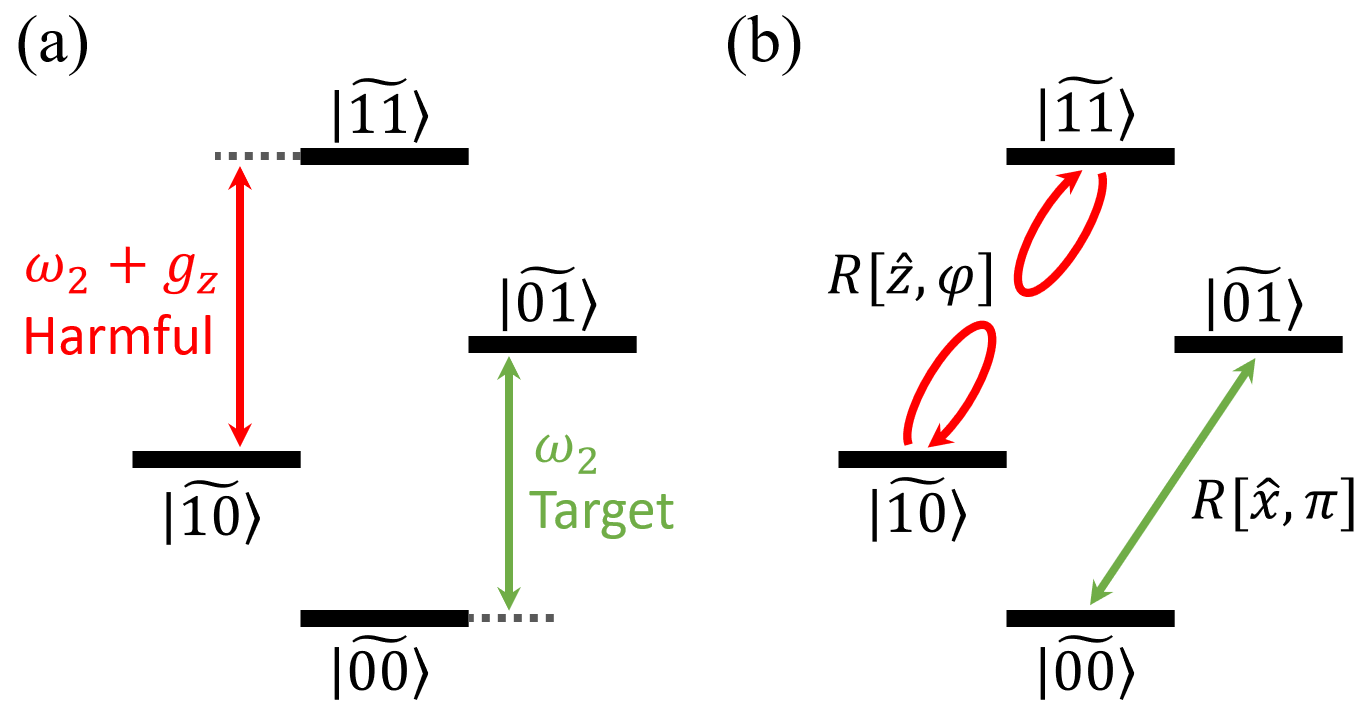}
    \end{center}
	\caption[Mechanics of a SWIPHT CNOT gate]{(a) The computational subspace of the two transmon qubit system with strong ZZ coupling. The frequency of the target transition (green) $\widetilde{\ket{00}}\leftrightarrow\widetilde{\ket{01}}$ is $\omega_{2}$. The harmful transition (red) $\widetilde{\ket{10}}\leftrightarrow\widetilde{\ket{11}}$ is detuned from the target transition by $g_z$. (b) The SWIPHT pulse (see \reffig{SWIPHT_CNOT_gate}) is on resonance with the target transition and induces a $\pi$ rotation around the $x$-axis in the two-level target subspace. At the same time, the harmful transition is off-resonantly driven by the SWIPHT pulse and undergoes cyclic evolution, acquiring trivial phases.}
\label{SWIPHT_CNOT_gate_energy_lvs}
\end{figure}

Fig.~\ref{SWIPHT_CNOT_gate_energy_lvs}(a) shows the energy levels in the computational subspace of the two-qubit system. A CNOT gate can be viewed as effectively swapping the $\widetilde{\ket{00}}$ and $\widetilde{\ket{01}}$ states while leaving the other logical states alone. A natural way to implement this is to apply a resonant microwave drive to perform a $\pi$ rotation on the target transition $\widetilde{\ket{00}}\leftrightarrow\widetilde{\ket{01}}$. However, there exists a nearby harmful transition $\widetilde{\ket{10}}\leftrightarrow\widetilde{\ket{11}}$ [red in \reffig{SWIPHT_CNOT_gate_energy_lvs}(a)] that is detuned from the target transition by $g_z$, so it would be driven off-resonantly. As a result, one would have to make the gate time much longer than $2\pi/g_z$ to selectively drive the target transition. This is not optimal, given the limited coherence times of the qubits. Instead of attempting to avoid the harmful transition, the SWIPHT CNOT gate is designed to purposely drive it through a trivial cyclic evolution while driving a $\pi$ rotation on the target transition. This is shown in \reffig{SWIPHT_CNOT_gate_energy_lvs}(b). A cyclic evolution on a two-level system can be described by a Z rotation, $e^{-i\varphi\sigma_{z}^{(2)}/2}$, where $\varphi$ is the phase accumulated during the cyclic evolution. We can express the SWIPHT CNOT gate in the dressed two-qubit basis as
\begin{equation}
\begin{split}
  U_{\text{CNOT}}&=\widetilde{\ket{0}}\widetilde{\bra{0}}\otimes e^{-i\pi\sigma_{x}^{(2)}/2}+\widetilde{\ket{1}}\widetilde{\bra{1}}\otimes e^{-i\varphi\sigma_{z}^{(2)}/2}\\
  &=
\begingroup 
\setlength\arraycolsep{8pt}
\renewcommand{\arraystretch}{1}
    \begin{pmatrix}
    0&-i&0&0\\
    -i&0&0&0\\
    0&0&e^{-i\frac{\varphi}{2}}&0\\
    0&0&0&e^{i\frac{\varphi}{2}}
    \end{pmatrix}.
\endgroup 
\label{CNOT_gate}
\end{split}
\end{equation}

Compared to a standard CNOT gate, the SWIPHT CNOT gate described above has some extra phases, which is why we refer to it as a generalized CNOT gate~\cite{economou2015analytical}. This generalized CNOT gate is maximally entangling and is related to the standard CNOT by single-qubit Z rotations and a global phase.

According to the SWIPHT protocol~\cite{economou2015analytical}, analytical pulse shapes that implement the operation given in \refeq{CNOT_gate} can be generated from the formula
\begin{equation}
    \Omega(t)=\frac{\ddot{\chi}}{\sqrt{\frac{g_{z}^{2}}{4}-\dot{\chi}^{2}}}-2\sqrt{\frac{g_{z}^{2}}{4}-\dot{\chi}^{2}}\cot{(2\chi)}.
    \label{OmegaSWIPHT}
\end{equation}
Any real function $\chi(t)$ obeying the constraint $|\dot\chi|\le g_z/2$ and the boundary conditions $\chi(0)=\chi(t_g)=\pi/4$, $\dot\chi(0)=\dot\chi(t_g)=0$, where $t_g$ is the gate time, produces a pulse wave form $\Omega(t)$ that performs a generalized CNOT.
One may notice a factor of two difference from the original expression in Ref.~\cite{economou2015analytical}; This is due to different definitions of the Rabi strength. Here, we choose $\chi(t)$ as in Ref.~\cite{economou2015analytical}:
\begin{equation}
    \chi(t)=A\left(\frac{t}{t_{g}}\right)^{4}\left(1-\frac{t}{t_{g}}\right)^4+\frac{\pi}{4},
\end{equation}
\begin{figure}[!htb]
    \begin{center}
	\includegraphics[width=8cm]{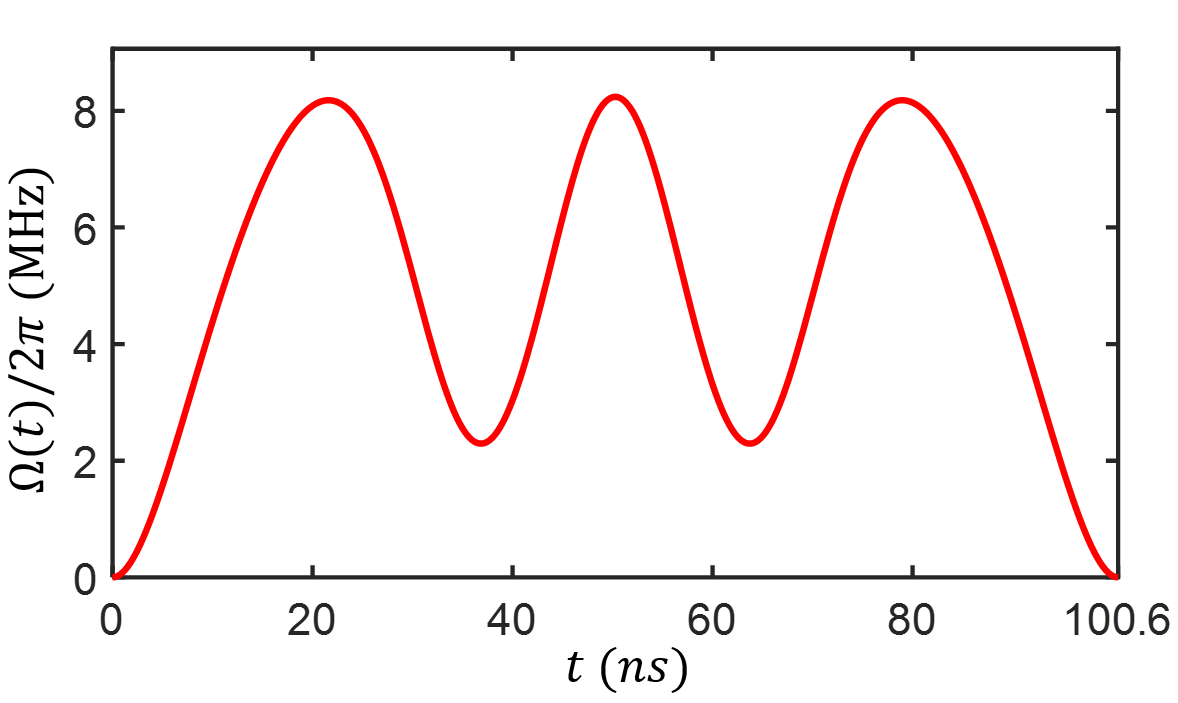}
    \end{center}
	\caption[Pulse shape that implements the SWIPHT CNOT gate]{Pulse shape that implements the SWIPHT CNOT gate. The gate time is about $\SI{100.6}{\ns}$.}
\label{SWIPHT_CNOT_gate}
\end{figure}
where $A$ and $t_g$ are given by
\begin{equation}
    t_{g}=\frac{5.87}{g_z},\quad A=138.9.
\end{equation}
Given $g_z=9.29$~MHz, the resulted gate time is $t_{g}=100.6$~ns. The pulse shape of the SWIPHT gate is plotted in~\reffig{SWIPHT_CNOT_gate}. The performance of our CNOT gate is evaluated in Secs.~\ref{sec:qst} and \ref{sec:qpt} through quantum state and process tomography.

\section{Randomized Benchmarking of the two-axis, single-qubit gate}\label{sec:randomizedbenchmarking}
To characterize the TAG, we perform the randomized benchmarking (RB) protocol described in Ref.~\cite{knill2008randomized} with assumption that errors are gate independent. The RB experiment starts with the qubit initialized in the ground state. Various $\pi/2$ gates and $\pi$ gates are randomly chosen and concatenated together to form a sequence. A comparison of the measured result of these operations with what would be expected from an ideal system can then be used to evaluate performance.

The $\pi/2$ gates are Clifford group generators $e^{\pm i\sigma_{u}\pi/4}$, with $u=x,y$. The $\pi$ gates are chosen from $e^{\pm i \sigma_{b}\pi/2}$ with $b=0,x,y,z$, where we have defined $\sigma_{0}$ to be the identity operator. The RB sequence is truncated at different numbers of $\pi/2-\pi$ gate pairs and applied to the qubit. Each truncation is followed by a $\pi/2$ or $\pi$ gate to always bring the qubit to a certain eigenstate of $\sigma_z$. For convenience, we typically use the ground state in our experiments. Then, the probability of the ground state, alternatively referred to as the sequence fidelity, is measured for each truncation of the RB sequence. The above process is repeated for many different RB sequences. Finally, the sequence fidelity $\mathcal{F}$ is averaged over all of the sequences, thereby becoming a function of only the truncation, which is quantified by the number of $\pi/2$ gates. $\mathcal{F}$ decays exponentially, and the decay rate is determined by the average gate infidelity~\cite{knill2008randomized}.
\begin{figure}[!ht]
	\includegraphics[width=8cm]{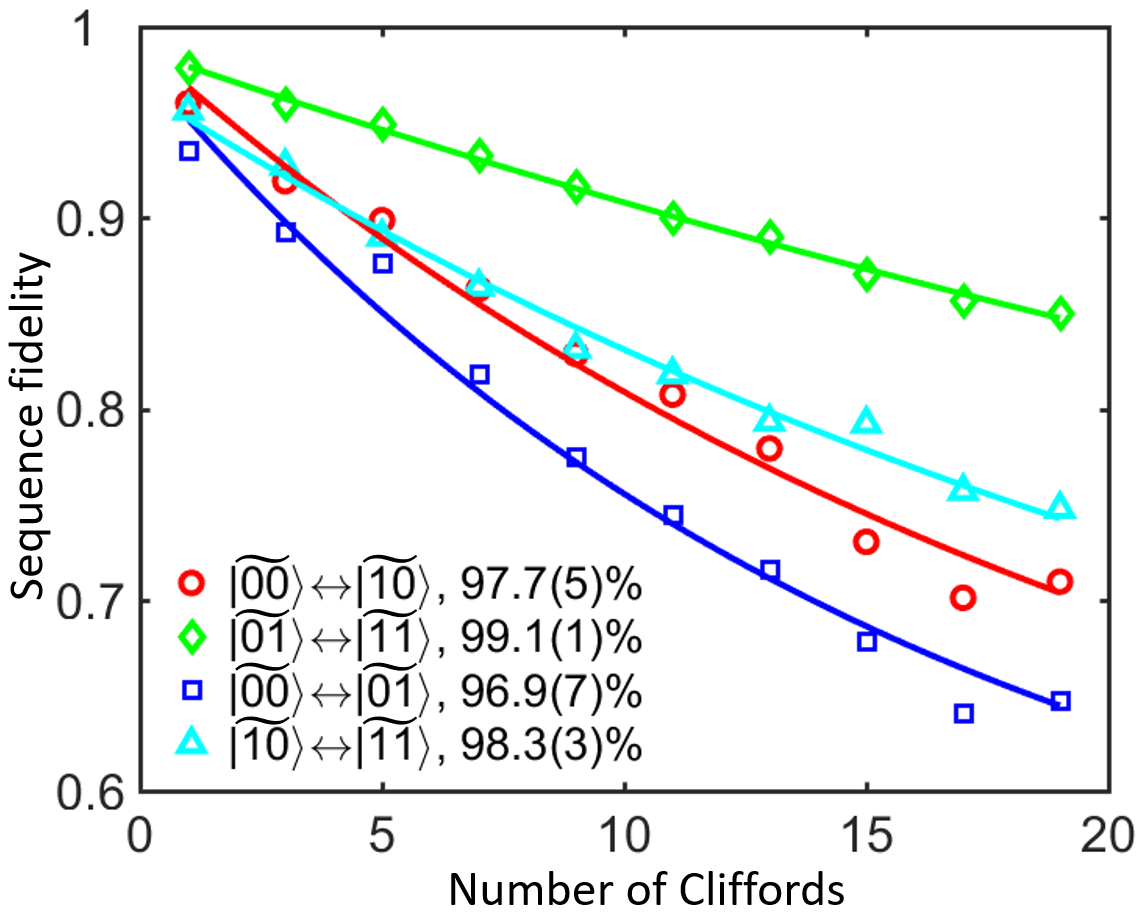}
	\caption[Randomized benchmarking of the two-axis gate]{Randomized benchmarking of the two-axis, single-qubit gate. Average sequency fidelity vs. the number of Cliffords/$\frac{\pi}{2}$ gates for the transition $\widetilde{\ket{00}}\leftrightarrow\widetilde{\ket{10}}$ (red circle), $\widetilde{\ket{01}}\leftrightarrow\widetilde{\ket{11}}$ (green diamond), $\widetilde{\ket{00}}\leftrightarrow\widetilde{\ket{01}}$ (blue square), and $\widetilde{\ket{10}}\leftrightarrow\widetilde{\ket{11}}$ (cyan triangle). Solid curves are the exponential fittings to the data to extract an average gate fidelity for each transition.}
\label{RB}
\end{figure}
Since the TAG is designed to address two transitions for each qubit (one where the other qubit is in the ground state, and one where it is in the excited state), we run the RB experiment on each transition separately for each qubit. This measures the average fidelity of the TAG as it acts on the corresponding transition. For each RB experiment, we choose 5 random $\pi/2$ gate sequences and 8 random $\pi$ gate sequences, yielding a total of 40 different RB sequences. Each sequence is truncated at every other pair of $\pi$ and $\pi/2$ gates. For each truncation, 1000 copies are measured to obtain the sequence fidelity $\mathcal{F}$. The experimental results are plotted in \reffig{RB} for all four transitions of the two qubits. The average gate fidelities of the TAG are $97.7(5)$\%, $99.1(1)$\%, $96.9(7)$\%, and $98.3(3)$\% for the transitions $\widetilde{\ket{00}}\leftrightarrow\widetilde{\ket{10}}$, $\widetilde{\ket{01}}\leftrightarrow\widetilde{\ket{11}}$, $\widetilde{\ket{00}}\leftrightarrow\widetilde{\ket{01}}$, and $\widetilde{\ket{10}}\leftrightarrow\widetilde{\ket{11}}$, respectively. Since each TAG has to be calibrated simultaneously for two transitions of a qubit, we believe there are some coherence errors in the TAGs due to imperfect calibration. The gate fidelities of the two transitions of the second qubit are lower than those of the first qubit. This is consistent with the fact that $T_{2}^{*}$ of the second qubit is only half that of the first qubit.

\section{Quantum State Tomography of a Maximally Entangled State}\label{sec:qst}
The core functionality of two-qubit gates is to entangle two qubits. We first verify this functionality by preparing a maximally entangled state and performing quantum state tomography on it. Since the CNOT gate is more straightforward in terms of generating entangled states, we present experimental results of an entangled state generated with the SWIPHT CNOT gate. 
\begin{figure}[!htb]
	\includegraphics[width=8.5cm]{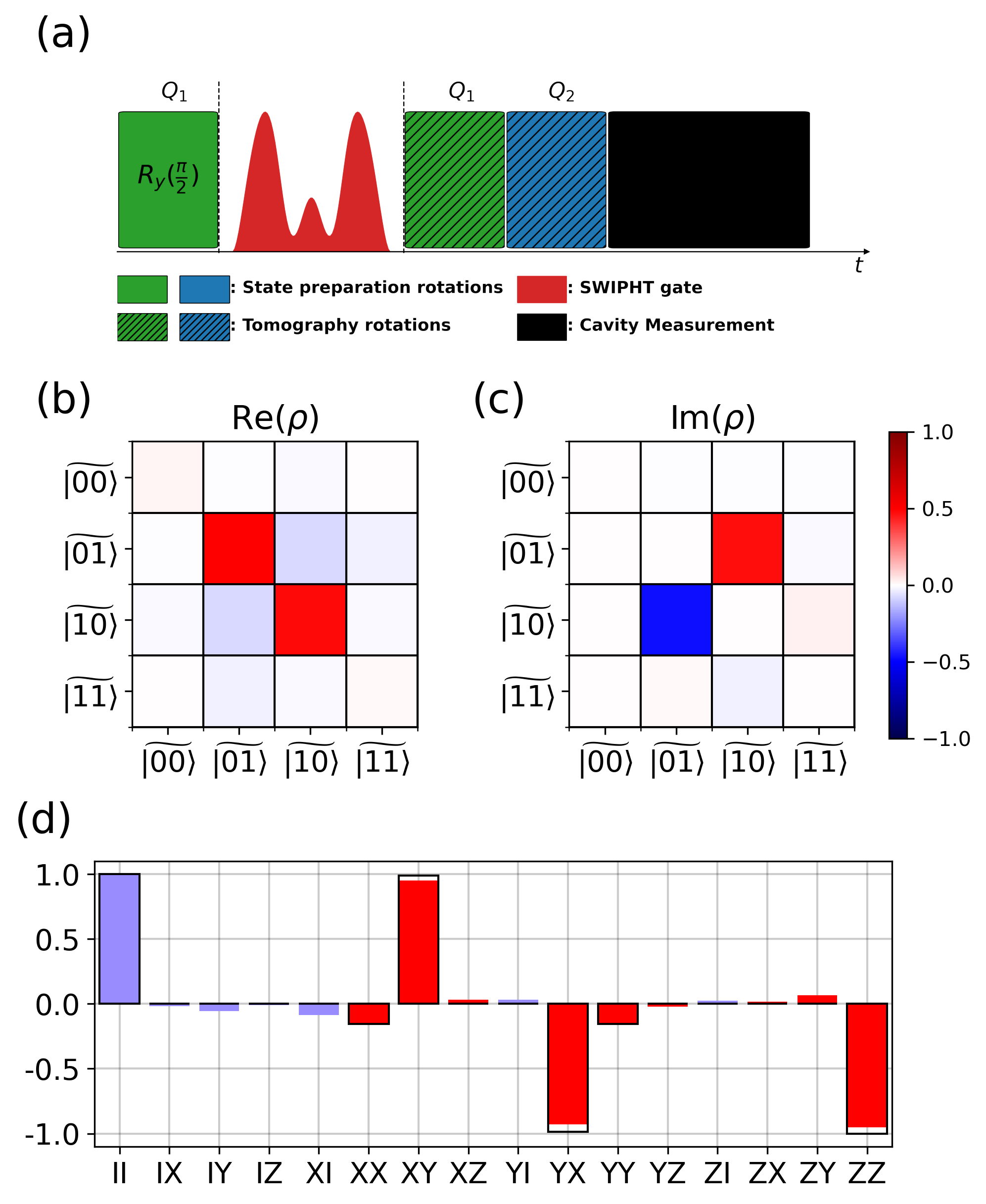}
	\caption[Results of quantum state tomography]{Results of quantum state tomography on a maximally entangled state prepared with the SWIPHT CNOT gate. (a) Pulse sequence for the quantum state tomography experiment (not to scale). (b) and (c) are the real and imaginary parts of the reconstructed density matrix. (d) The vector representation of the state in the Pauli basis. The light blue bars are the single-qubit components, and the red bars are the two-qubit components. The outlined bars represent the ideal entangled state $\frac{1}{\sqrt{2}}(e^{-i\frac{\varphi}{2}}\ket{10}-i\ket{01})$. By comparing to the ideal entangled state, we obtained an experimental fidelity of 98.2\%.}
\label{QST_results}
\end{figure}

As shown in \reffig{QST_results}(a), the experiment starts from a $\pi/2$ rotation around the $y$-axis of the first qubit, which is the control qubit. This prepares the system in the state $\frac{1}{\sqrt{2}}(\ket{0}+\ket{1})\ket{0}$. We then apply the SWIPHT CNOT gate, which brings the system to a maximally entangled state $\ket{\psi} = \frac{1}{\sqrt{2}}(e^{-i\frac{\varphi}{2}}\ket{10}-i\ket{01})$. The extra phase $\varphi/2$ on $\ket{10}$ is due to the fact that the SWIPHT CNOT is a generalized CNOT gate. Finally, tomography rotations are applied, followed by projective measurements on both qubits. Each run consists of $N=1000$ measurements in order to obtain good statistics on the final state populations. Noise and imperfections during the experiment can lead to an unphysical density matrix. To remedy this, a maximum likelihood estimation is applied to search for a physical density matrix such that the predicted probability of obtaining the observed data with the physical density matrix is maximized~\cite{o2004quantum}. In addition, a wide-band, quantum limited Josephson parametric amplifier was used that allowed us to read out both qubits at the same time. 

The real and imaginary parts of the reconstructed density matrix are shown in \reffig{QST_results}(b) and (c), respectively. As we can see, the largest contributions are all within the $\ket{10}$, $\ket{01}$ subspace, as expected. To compare to the ideal entangled state, the vector representation in the Pauli basis of the entangled state is shown in \reffig{QST_results}(d). A fidelity of 98.2~\% is obtained using $\mathcal{F}=\ev{\rho}{\psi}$, where $\ket{\psi}$ and $\rho$ are the ideal entangled state and the reconstructed density matrix from experiment, respectively.

\section{Quantum Process Tomography of Two-Qubit Gates}\label{sec:qpt}
Finally, we perform a full characterization of our two-qubit entangling gates using quantum process tomography. The process tomography starts from single-qubit rotations implemented with TAGs on both qubits to prepare the system. There are 36 different possible initial states for the two qubits: $\{\ket{\pm{}x}, \ket{\pm{}y}, \ket{\pm{}z}\}^{\otimes2}$. A separate run for each initial state is conducted. In each run, a two-qubit entangling gate is applied after the initial state is prepared. For the SWIPHT CNOT gate, the gate time is about 100.6~ns. However, to remove a residual conditional phase due to the ZZ interaction, right after the SWIPHT gate we let the system idle (evolve freely) for a time such that the sum of the SWIPHT gate time and the idle time is equal to a full period of the ZZ interaction, $2\pi/g_z$. For the free-evolution CZ gate, no entangling microwave pulse is applied, and the system instead undergoes free evolution for 53.8~ns. Subsequently, tomography rotations and projective measurements are performed on both qubits. Each run consists of $N=1000$ measurements in order to obtain good statistics on the final state populations. We apply a maximum likelihood estimation method to reconstruct the Pauli transfer matrix $R$ for the process. 

\begin{figure}[!htb]
    \begin{center}
	\includegraphics[width=9cm]{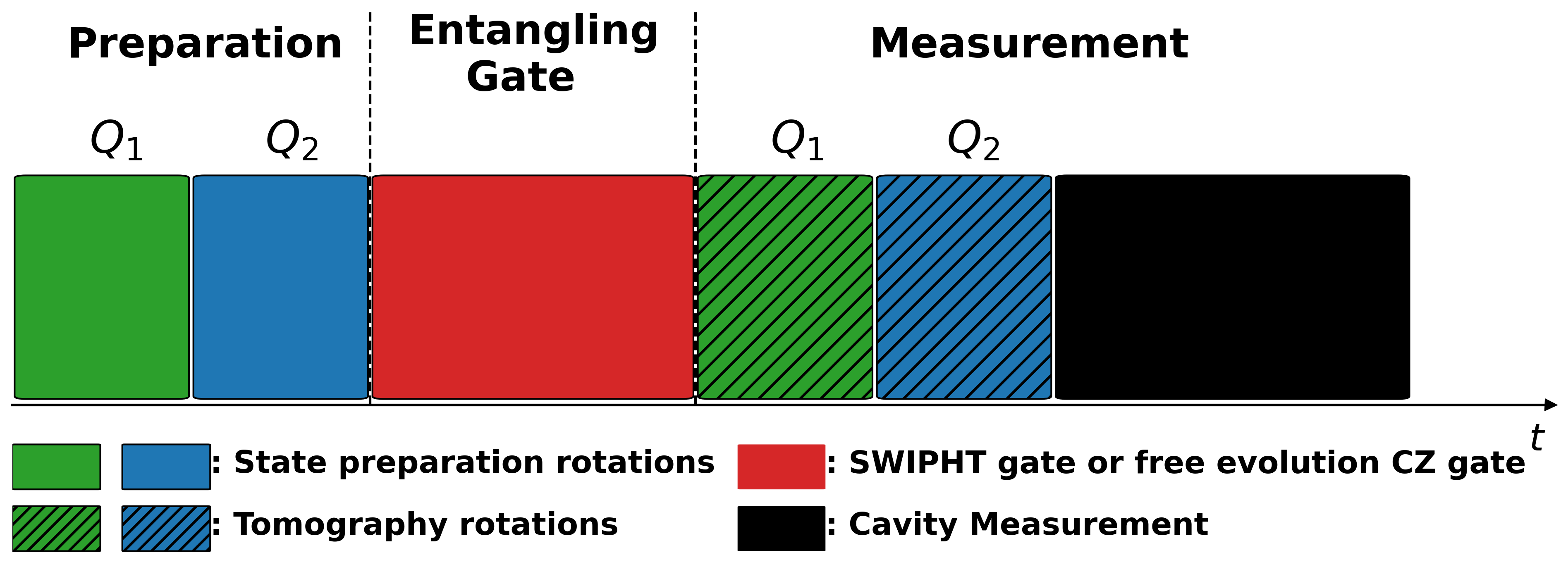}
    \end{center}
	\caption[Quantum process tomography of the SWIPHT CNOT gate and the free evolution CZ gate]{Pulse sequence for the quantum process tomography experiment (not to scale).}
\label{QPT_all_FlowChart}
\end{figure}

\begin{figure*}[!htb]
    \begin{center}
	\includegraphics[width=18cm]{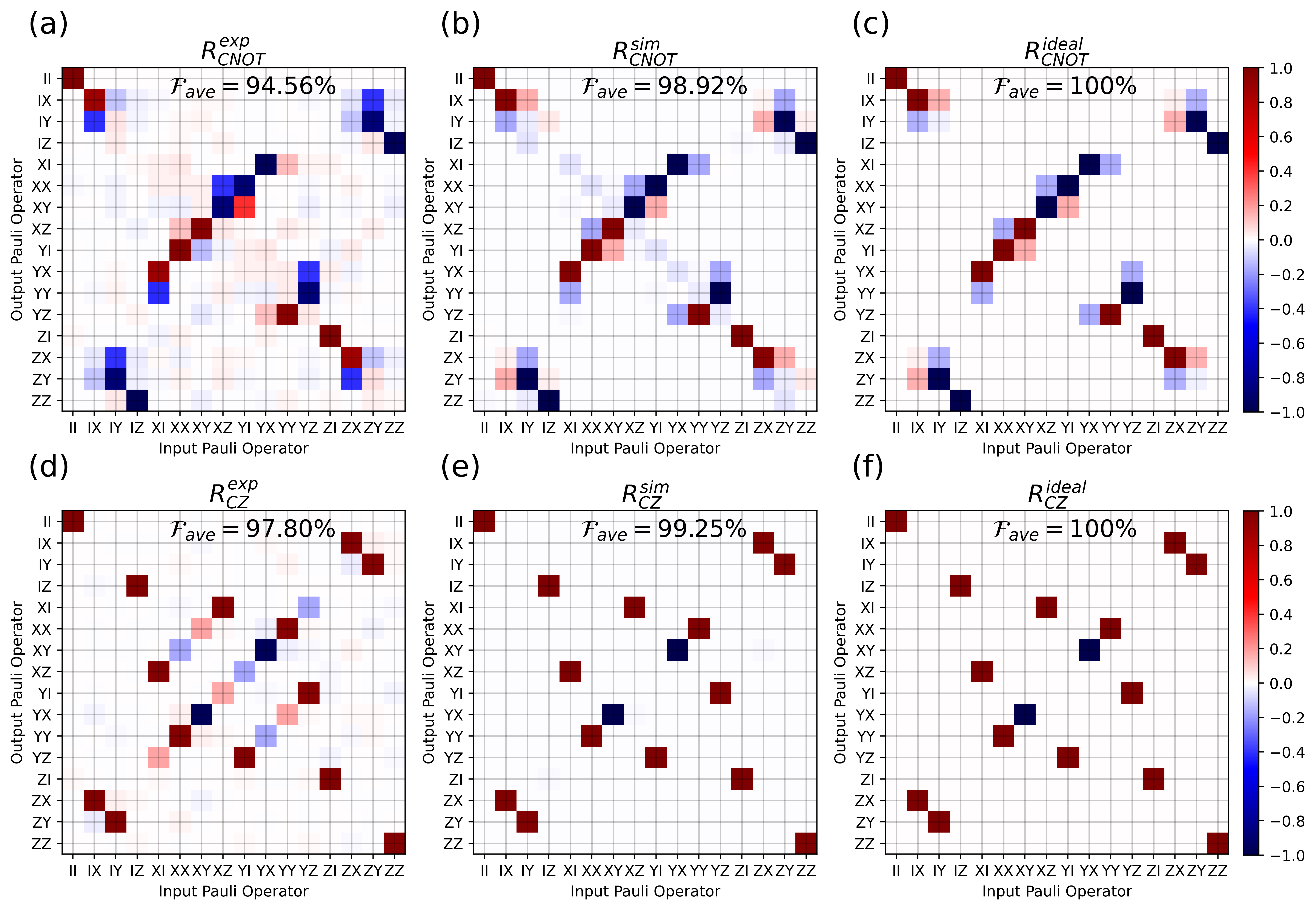}
    \end{center}
	\caption[Results of the quantum process tomography]{(a)-(c) [(d)-(f)] are the experimental, simulated, ideal Pauli transfer matrices for the SWIPHT CNOT gate [free evolution CZ gate]. Note that (e) and (f) are different although no significant difference can be noticed under this color scale.}
\label{QPT_results}
\end{figure*}

The experimental Pauli transfer matrices, $R_{\text{CNOT}}^{\text{exp}}$ for the SWIPHT CNOT gate and $R_{\text{CZ}}^{\text{exp}}$ for the free-evolution CZ gate, are shown in \reffig{QPT_results}(a) and (d), respectively. Figure~\ref{QPT_results}(c) and (f) illustrate the ideal $R$ matrices, $R_{\text{CNOT}}^{\text{ideal}}$ and $R_{\text{CNOT}}^{\text{ideal}}$, for the SWIPHT CNOT and CZ gate, respectively. The average gate fidelity in each case is computed by comparing $R^{\text{exp}}$ to $R^{\text{ideal}}$,
\begin{equation}
    \mathcal{F}_{ave}=\frac{\text{Tr}[(R^{\text{exp}})^{T}R^{\text{ideal}}]/4+1}{5}.
\end{equation}
For the SWIPHT CNOT (CZ), we obtain an average gate fidelity of 94.6\% (97.8\%). To understand how much the decoherence contributes to the infidelity of the gates, we performed master equation simulations for both the SWIPHT CNOT and CZ gates with state preparation rotations also included, incorporating decoherence at levels consistent with our measured $T_1$ and $T_2^*$ times. We then run the quantum process tomography on the simulated data to extract the simulated $R$ matrices, shown in \reffig{QPT_results}(b) and (e) for the SWIPHT CNOT and CZ gates, respectively. The average gate fidelities from the master equation simulations are 98.92\% for the SWIPHT CNOT gate and 99.25\% for the CZ gate. In another simulation, perfect state preparation is assumed. The outcome gate fidelities are 99.4\% for the SWIPHT CNOT gate and 99.7\% for the CZ gate. This indicates that the SPAM errors~\cite{dewes2012characterization} contributes about 0.5\% to the infidelity.
The experiments report about 4\% lower fidelities compared to the simulations. We attribute this difference to coherent errors~\cite{kaye2007introduction} and leakage errors~\cite{mckay2016universal}. Note that our master equation simulation only includes the computational subspace and therefore does not capture leakage errors. Purity of the SWIPHT CNOT Pauli transfer matrix is measured to be 95.54\%, which indicates the presence of leakage errors. Extra nonzero terms in \reffig{QPT_results}(a) and (d) compared to the simulation results in \reffig{QPT_results}(b) and (e) indicates that there are indeed coherent errors. On the other hand, the maximally entangled state in the quantum state tomography described in the last section reports a fidelity of 98.2\%, which is close to the decoherence limit from the master equation simulation. This means that other states in the quantum process tomography would have fidelities lower than the average gate fidelity. The varying state fidelities of different states in the quantum process tomography also indicates that there are coherent errors in the experiment.

\section{Conclusions}\label{sec:conclusions}
In this paper, we experimentally demonstrated a universal quantum gate set for the regime of strong ZZ coupling between superconducting transmon qubits. We introduced the concept of two-axis gates to implement arbitrary single-qubit rotations in this regime, and we tested their performance in a two-qubit system using randomized benchmarking. The average gate fidelity was found to be about 99\% for the first qubit and 98\% for the second qubit. In addition, we demonstrated two types of two-qubit entangling gates: a SWIPHT CNOT gate implemented with a shaped microwave pulse and a CZ gate based on free evolution under the ZZ interaction. The SWIPHT CNOT was found to produce a maximally entangled state with a measured fidelity of 98.2\% from quantum state tomography. We further characterized the performance of both the SWIPHT CNOT and the free-evolution CZ gates using quantum process tomography. We obtained average gate fidelities of 94.6\% and 97.8\% for the SWIPHT CNOT and the CZ, respectively. We presented evidence that the infidelities are likely due to coherent, SPAM, and leakage errors. Advanced pulse shaping techniques~\cite{motzoi2009simple} and calibration improvements would likely reduce these errors significantly. The universal gate set demonstrated here offers additional control flexibility that can help to optimize the use of tunable couplers and to mitigate the adverse effects of residual ZZ interactions in superconducting qubit processors.

\section{acknowledgments}
This research was supported by the National Science Foundation (Award No. 1839136), DOE (Grant No. DE-SC0019199) and the DOE through FNAL as well as the new and emerging qubit science and technology (NEQST) program initiated by the US Army
Research Office (ARO) in collaboration with the Laboratory of Physical Sciences (LPS) under Grant No. W911NF-18-1-0114. NIST authors acknowledge support of the NIST Quantum Based Metrology Initiative, NQI, and partial support from Google. This work is property of the US Government and not subject to copyright.

Junling Long and Tongyu Zhao contributed equally to this work.

\appendix
\section{Spectrum of the two-qubit system measured with Ramsey oscillations}\label{app:spectrum}
\begin{table}[!htb]
    \centering
    \def\arraystretch{1.5}
    \begin{tabular}{|c|c|}
        \hline
        Qubit Transition & Frequency(\si{\GHz}) \\
        \hline
        $\widetilde{\ket{00}}\leftrightarrow\widetilde{\ket{10}}$ & 5.07478658\\
        \hline
        $\widetilde{\ket{00}}\leftrightarrow\widetilde{\ket{01}}$ & 5.30990762\\
        \hline
        $\widetilde{\ket{01}}\leftrightarrow\widetilde{\ket{11}}$ & 5.08406906\\
        \hline
        $\widetilde{\ket{10}}\leftrightarrow\widetilde{\ket{11}}$ & 5.31920716\\
        \hline
        $\widetilde{\ket{10}}\leftrightarrow\widetilde{\ket{20}}$ & 4.81503094\\
        \hline
        $\widetilde{\ket{10}}\leftrightarrow\widetilde{\ket{02}}$ & 4.96857665\\
        \hline
    \end{tabular}
    \caption{Spectrum of the two-qubit system.}
    \label{tab:spectrum}
\end{table}

The spectrum of the coupled two-qubit transmon system is shown in Table~\ref{tab:spectrum}.

\section{Relaxation and Ramsey times}\label{app:decoherence}
\begin{table}[!htb]
    \centering
    \begin{tabular}{|c|c|c|}
        \hline
         & $T_1(\si{\us})$ & $T_2^*(\si{\us})$ \\
        \hline
        $Q_1$ & 76.98 & 50.65\\
        \hline
        $Q_2$ & 79.71 & 17.09\\
        \hline
    \end{tabular}
    \caption{Relaxation and Ramsey times of the qubits.}
    \label{tab:coherence}
\end{table}

The Relaxation and Ramsey times of the coupled two-qubit system are shown in Table~\ref{tab:coherence}.

\section{Derivation of the ZZ coupling}\label{app:ZZcoupling}

In this appendix, we derive effective ZZ-coupling Hamiltonians for two coupled transmons. We first consider the case of a direct capacitive coupling, and then we consider an indirect, cavity-mediated coupling. In both cases, an effective Hamiltonian of the form shown in Eq.~\ref{2QZZHamiltonian} is obtained.  Similar derivations have been given in prior works \cite{Koch2007,mckay2019three}, but we include them here for the sake of completeness.

\subsection{Direct capacitive coupling}

Two transmons that have a direct capacitive coupling are described by the Hamiltonian
\bea
H&=&\sum_j\omega_{1,j}\ket{j}\bra{j}+\sum_{\alpha}\omega_{2,\alpha}\ket{\alpha}\bra{\alpha}\nonumber\\&+&\sum_{j\alpha}g_{j,\alpha}\ket{j}\bra{j+1}\otimes\ket{\alpha+1}\bra{\alpha}+\hbox{H.c.}\label{eq:capacitiveHam}
\eea
The energy of the $j$th level of the first transmon is $\omega_{1,j}$, while the energy of the $\alpha$th level of the second transmon is $\omega_{2,\alpha}$. We use Latin indices to index transmon 1 states and Greek indices to index transmon 2 states. The coupling strengths between these levels are denoted by $g_{j,\alpha}$. We assume these are real.

We will perform a Schrieffer-Wolff transformation to eliminate the coupling terms in Eq.~\eqref{eq:capacitiveHam}. This will be done using a similarity transformation $D=e^{S-S^\dagger}$, where
\be
S=\sum_{j\alpha}\beta_{j,\alpha}\ket{j+1}\bra{j}\otimes\ket{\alpha}\bra{\alpha+1}.
\ee
Under this transformation, the Hamiltonian becomes
\be
H'=DHD^\dagger=H+[S-S^\dagger,H]+\frac{1}{2}[S-S^\dagger,[S-S^\dagger,H]]+\ldots
\ee
Our goal is to find the values of $\beta_{j,\alpha}$ that make the terms in $[S-S^\dagger,H]$ that are linear in $g_{j,\alpha}$ cancel the coupling terms in $H$.

For each Hermitian term $A$ in Eq.~\eqref{eq:capacitiveHam}, it suffices to compute $[S,A]$, because $[S-S^\dagger,A]=[S,A]+[S,A]^\dagger$. Keeping only terms up to second order in $g_{j,\alpha}$ and $\beta_{j,\alpha}$, we have
\bea\label{eq:commSwithQ1}
&&\left[S,\sum_j\omega_{1,j}\ket{j}\bra{j}\right]\nonumber\\&&=\sum_{j\alpha}\beta_{j,\alpha}(\omega_{1,j}-\omega_{1,j+1})\ket{j+1}\bra{j}\otimes\ket{\alpha}\bra{\alpha+1},\nonumber\\
\eea
\bea\label{eq:commSwithQ2}
&&\left[S,\sum_\alpha\omega_{2,\alpha}\ket{\alpha}\bra{\alpha}\right]\nonumber\\&&=\sum_{j\alpha}\beta_{j,\alpha}(\omega_{2,\alpha+1}-\omega_{2,\alpha})\ket{j+1}\bra{j}\otimes\ket{\alpha}\bra{\alpha+1},\nonumber\\
\eea
\bea\label{eq:commSwithCoupling1}
&&\left[S,\sum_{j\alpha}g_{j,\alpha}\ket{j}\bra{j+1}\otimes\ket{\alpha+1}\bra{\alpha}\right]\nonumber\\&&=\sum_{j\alpha}\beta_{j,\alpha}g_{j,\alpha}\ket{j+1}\bra{j+1}\otimes\ket{\alpha}\bra{\alpha}\nonumber\\&&-\sum_{j\alpha}\beta_{j,\alpha}g_{j,\alpha}\ket{j}\bra{j}\otimes\ket{\alpha+1}\bra{\alpha+1},\nonumber\\
\eea
\bea\label{eq:commSwithCoupling2}
&&\left[S,\sum_{j\alpha}g_{j,\alpha}\ket{j+1}\bra{j}\otimes\ket{\alpha}\bra{\alpha+1}\right]\nonumber\\&&=\sum_{j\alpha}\beta_{j+1,\alpha}g_{j,\alpha+1}\ket{j+2}\bra{j}\otimes\ket{\alpha}\bra{\alpha+2}\nonumber\\&&-\sum_{j\alpha}\beta_{j,\alpha+1}g_{j+1,\alpha}\ket{j+2}\bra{j}\otimes\ket{\alpha}\bra{\alpha+2},\nonumber\\
\eea
\bea
&&\left[S,\left[S-S^\dagger,\sum_j\omega_{1,j}\ket{j}\bra{j}\right]\right]\nonumber\\&&=\sum_{j\alpha}\beta_{j,\alpha}^2(\omega_{1,j}-\omega_{1,j+1})\ket{j+1}\bra{j+1}\otimes\ket{\alpha}\bra{\alpha}\nonumber\\&&-\sum_{j\alpha}\beta_{j,\alpha}^2(\omega_{1,j}-\omega_{1,j+1})\ket{j}\bra{j}\otimes\ket{\alpha+1}\bra{\alpha+1}\nonumber\\&&+\sum_{j\alpha}\beta_{j+1,\alpha}\beta_{j,\alpha+1}(\omega_{1,j}-\omega_{1,j+1})\ket{j+2}\bra{j}\otimes\ket{\alpha}\bra{\alpha+2}\nonumber\\&&-\sum_{j\alpha}\beta_{j,\alpha+1}\beta_{j+1,\alpha}(\omega_{1,j+1}-\omega_{1,j+2})\ket{j+2}\bra{j}\otimes\ket{\alpha}\bra{\alpha+2},\nonumber\\
\eea
\bea
&&\left[S,\left[S-S^\dagger,\sum_\alpha\omega_{2,\alpha}\ket{\alpha}\bra{\alpha}\right]\right]\nonumber\\&&=\sum_{j\alpha}\beta_{j,\alpha}^2(\omega_{2,\alpha+1}-\omega_{2,\alpha})\ket{j+1}\bra{j+1}\otimes\ket{\alpha}\bra{\alpha}\nonumber\\&&-\sum_{j\alpha}\beta_{j,\alpha}^2(\omega_{2,\alpha+1}-\omega_{2,\alpha})\ket{j}\bra{j}\otimes\ket{\alpha+1}\bra{\alpha+1}\nonumber\\&+&\sum_{j\alpha}\beta_{j+1,\alpha}\beta_{j,\alpha+1}(\omega_{2,\alpha+2}-\omega_{2,\alpha+1})\ket{j+2}\bra{j}\otimes\ket{\alpha}\bra{\alpha+2}\nonumber\\&&-\sum_{j\alpha}\beta_{j,\alpha+1}\beta_{j+1,\alpha}(\omega_{2,\alpha+1}-\omega_{2,\alpha})\ket{j+2}\bra{j}\otimes\ket{\alpha}\bra{\alpha+2},\nonumber\\
\eea
From Eqs.~\eqref{eq:commSwithQ1} and \eqref{eq:commSwithQ2}, we see that we need 
\be
\beta_{j,\alpha}=-\frac{g_{j,\alpha}}{\omega_{1,j}-\omega_{1,j+1}+\omega_{2,\alpha+1}-\omega_{2,\alpha}}.
\ee
Eq.~\eqref{eq:commSwithCoupling1} then gives rise to shifts in the transmon energy levels and to an effective ZZ coupling. Restricting to the logical subspace, the full Hamiltonian then becomes
\bea
H'_{logical}&=&\frac{1}{2}\left(\omega_{1,0}+\omega_{1,1}+\omega_{2,0}+\omega_{2,1}+\frac{1}{2}\lambda_{ZZ}\right)II\nonumber\\&+&\frac{1}{2}\left(\omega_{2,0}-\omega_{2,1}-\eta-\frac{1}{2}\lambda_{ZZ}\right)IZ\nonumber\\
&+&\frac{1}{2}\left(\omega_{1,0}-\omega_{1,1}+\eta-\frac{1}{2}\lambda_{ZZ}\right)ZI+\frac{1}{4}\lambda_{ZZ}ZZ,\nonumber\\
\eea
where
\be
\eta=\frac{g_{0,0}^2}{\omega_{1,0}+\omega_{2,1}-\omega_{1,1}-\omega_{2,0}},
\ee
\be
\lambda_{ZZ}=\frac{g_{1,0}^2}{\omega_{1,1}+\omega_{2,1}-\omega_{1,2}-\omega_{2,0}}+\frac{g_{0,1}^2}{\omega_{1,1}+\omega_{2,1}-\omega_{1,0}-\omega_{2,2}}.
\ee
We can see that the frequencies of the transitions $\widetilde{\ket{00}}\leftrightarrow\widetilde{\ket{01}}$ and $\widetilde{\ket{10}}\leftrightarrow\widetilde{\ket{11}}$ in the dressed basis are split by $\lambda_{ZZ}$. Notice that $\lambda_{ZZ}$ itself depends quadratically on the bare couplings and inversely on the bare energy splittings between $\ket{11}$ and $\ket{20}$ and between $\ket{11}$ and $\ket{02}$. If we define $\omega_1=\omega_{1,1}-\omega_{1,0}-\eta$, $\omega_2=\omega_{2,1}-\omega_{2,0}+\eta$, and $g_{z}=\lambda_{ZZ}$, we obtain Eq.~\ref{2QZZHamiltonian}.

\subsubsection{Driving terms}

We now consider the inclusion of driving terms on each transmon:
\be
H_d=\Omega_1(t)\sum_{j}d_{1,j}\ket{j}\bra{j+1}+\Omega_2(t)\sum_{\alpha}d_{2,\alpha}\ket{\alpha}\bra{\alpha+1}+\hbox{H.c.}
\ee
Under the Schrieffer-Wolff transformation, this becomes
\be
H_d'=DH_dD^\dagger=H_d+[S-S^\dagger,H_d]+\frac{1}{2}[S-S^\dagger,[S-S^\dagger,H_d]]+\ldots
\ee
where
\begin{widetext}
\bea
\left[S,\Omega_1(t)\sum_{j}d_{1,j}\ket{j}\bra{j+1}\right]&=&\Omega_{1}(t)\sum_{j\alpha}\beta_{j,\alpha}d_{1,j}\ket{j+1}\bra{j+1}\otimes\ket{\alpha}\bra{\alpha+1}-\Omega_1(t)\sum_{j,\alpha}\beta_{j,\alpha}d_{1,j}\ket{j}\bra{j}\otimes\ket{\alpha}\bra{\alpha+1}\nonumber\\&=&\Omega_1(t)\sum_{j\alpha}\left(\beta_{j,\alpha}d_{1,j}-\beta_{j+1,\alpha}d_{1,j+1}\right)\ket{j+1}\bra{j+1}\otimes\ket{\alpha}\bra{\alpha+1}\nonumber\\&&-\;\Omega_1(t)d_{1,0}\ket{0}\bra{0}\otimes\sum_\alpha\beta_{0,\alpha}\ket{\alpha}\bra{\alpha+1}.
\eea
\bea
\left[S,\Omega_1^*(t)\sum_{j}d_{1,j}^*\ket{j+1}\bra{j}\right]&=&\Omega_1^*(t)\sum_{j\alpha}\beta_{j+1,\alpha}d_{1,j}^*\ket{j+2}\bra{j}\otimes\ket{\alpha}\bra{\alpha+1}\nonumber\\&&-\;\Omega_1^*(t)\sum_{j,\alpha}\beta_{j,\alpha}d_{1,j+1}^*\ket{j+2}\bra{j}\otimes\ket{\alpha}\bra{\alpha+1}\nonumber\\&=&\Omega_1^*(t)\sum_{j\alpha}\left(\beta_{j+1,\alpha}d_{1,j}^*-\beta_{j,\alpha}d_{1,j+1}^*\right)\ket{j+2}\bra{j}\otimes\ket{\alpha}\bra{\alpha+1}.
\eea
\bea
\left[S,\Omega_2(t)\sum_{\alpha}d_{2,\alpha}\ket{\alpha}\bra{\alpha+1}\right]&=&\Omega_2(t)\sum_{j\alpha}\beta_{j,\alpha}d_{2,\alpha+1}\ket{j+1}\bra{j}\otimes\ket{\alpha}\bra{\alpha+2}\nonumber\\&&-\;\Omega_2(t)\sum_{j,\alpha}\beta_{j,\alpha+1}d_{2,\alpha}\ket{j+1}\bra{j}\otimes\ket{\alpha}\bra{\alpha+2}\nonumber\\&=&\Omega_2(t)\sum_{j\alpha}\left(\beta_{j,\alpha}d_{2,\alpha+1}-\beta_{j,\alpha+1}d_{2,\alpha}\right)\ket{j+1}\bra{j}\otimes\ket{\alpha}\bra{\alpha+2}.
\eea
\bea
\left[S,\Omega_2^*(t)\sum_{\alpha}d_{2,\alpha}^*\ket{\alpha+1}\bra{\alpha}\right]&=&\Omega_2^*(t)\sum_{j\alpha}\beta_{j,\alpha}d_{2,\alpha}^*\ket{j+1}\bra{j}\otimes\ket{\alpha}\bra{\alpha}-\Omega_2^*(t)\sum_{j,\alpha}\beta_{j,\alpha}d_{2,\alpha}^*\ket{j+1}\bra{j}\otimes\ket{\alpha+1}\bra{\alpha+1}\nonumber\\&=&\Omega_2^*(t)\sum_{j\alpha}\left(\beta_{j,\alpha+1}d_{2,\alpha+1}^*-\beta_{j,\alpha}d_{2,\alpha}^*\right)\ket{j+1}\bra{j}\otimes\ket{\alpha+1}\bra{\alpha+1}\nonumber\\&&+\;\Omega_2^*(t)d_{2,0}^*\sum_j\beta_{j,0}\ket{j+1}\bra{j}\otimes\ket{0}\bra{0}.
\eea
In the logical subspace, we then have
\bea
[S-S^\dagger,H_d]_{logical}&=&\ket{1}\bra{1}\otimes\left[\Omega_1(\beta_{0,0}d_{1,0}-\beta_{1,0}d_{1,1})\ket{0}\bra{1}+\Omega_1^*(\beta_{0,0}d_{1,0}^*-\beta_{1,0}d_{1,1}^*)\ket{1}\bra{0}\right]\nonumber\\
&-&\beta_{0,0}\ket{0}\bra{0}\otimes\left[\Omega_1d_{1,0}\ket{0}\bra{1}+\Omega_1^*d_{1,0}^*\ket{1}\bra{0}\right]\nonumber\\
&+&\left[\Omega_2(\beta_{0,1}d_{2,1}-\beta_{0,0}d_{2,0})\ket{0}\bra{1}+\Omega_2^*(\beta_{0,1}d_{2,1}^*-\beta_{0,0}d_{2,0}^*)\ket{1}\bra{0}\right]\otimes\ket{1}\bra{1}\nonumber\\
&+&\beta_{0,0}\left[\Omega_2d_{2,0}\ket{0}\bra{1}+\Omega_2^*d_{2,0}^*\ket{1}\bra{0}\right]\otimes\ket{0}\bra{0}.
\eea
The full effective Hamiltonian in the logical subspace is\\\\
\begin{equation*}
\resizebox{1.0 \textwidth}{!}{$
H'_{logical}=
  \begin{pmatrix} 
    \omega_{1,0}+\omega_{2,0} & d_{2,0}\Omega_2-d_{1,0}\beta_{0,0}\Omega_1 & d_{1,0}\Omega_1+d_{2,0}\beta_{0,0}\Omega_2 & 0 \\
    d_{2,0}^*\Omega_2^*-d_{1,0}^*\beta_{0,0}\Omega_1^* & \omega_{1,0}+\omega_{2,1}+\eta & 0 & d_{1,0}\Omega_1+(d_{2,1}\beta_{0,1}-d_{2,0}\beta_{0,0})\Omega_2 \\
    d_{1,0}^*\Omega_1^*+d_{2,0}^*\beta_{0,0}\Omega_2^* & 0 & \omega_{1,1}+\omega_{2,0}-\eta & (d_{1,0}\beta_{0,0}-d_{1,1}\beta_{1,0})\Omega_1+d_{2,0}\Omega_2 \\
    0 & d_{1,0}^*\Omega_1^*+(d_{2,1}^*\beta_{0,1}-d_{2,0}^*\beta_{0,0})\Omega_2^* & (d_{1,0}^*\beta_{0,0}-d_{1,1}^*\beta_{1,0})\Omega_1^*+d_{2,0}^*\Omega_2^* & \omega_{1,1}+\omega_{2,1}+\lambda_{ZZ}
  \end{pmatrix}.
$}
\end{equation*}
\end{widetext}

\subsection{Resonator-mediated coupling}

Next, we consider two transmons coupled to the same resonator:
\bea
H&=&\omega_ca^\dagger a+\sum_j\epsilon_{1,j}\ket{j}\bra{j}+\sum_\alpha\epsilon_{2,\alpha}\ket{\alpha}\bra{\alpha}\\&&+\;\sum_jh_{1,j}a^\dagger\ket{j}\bra{j+1}+\sum_\alpha h_{2,\alpha} a^\dagger\ket{\alpha}\bra{\alpha+1}+\hbox{H.c.}\nonumber
\eea
$\omega_c$ is the cavity frequency.
The energy levels of the first transmon are $\epsilon_{1,j}$, while those of the second are $\epsilon_{2,\alpha}$.
$h_{1,j}$ and $h_{2,\alpha}$ are the transmon-cavity coupling strengths, which we assume are real. We again employ a Schrieffer-Wolff transformation $D=e^{S-S^\dagger}$ to transform to a new basis in which the Hamiltonian, $H'=DHD^\dagger$, does not have transmon-cavity couplings to first order in $h_{1,j}$ and $h_{2,\alpha}$. This is accomplished by choosing
\be
S=\sum_{j}\gamma_{1,j}a^\dagger\ket{j}\bra{j+1}+\sum_\alpha\gamma_{2,\alpha}a^\dagger\ket{\alpha}\bra{\alpha+1}.
\ee
To determine $\gamma_{1,j}$ and $\gamma_{2,\alpha}$, we compute commutators up to second order in the transmon-cavity couplings:
\bea\label{eq:commSwithCavity}
&&\left[S,\omega_ca^\dagger a\right]\nonumber\\&&=-\omega_c\sum_j\gamma_{1,j}a^\dagger\ket{j}\bra{j+1}-\omega_c\sum_\alpha\gamma_{2,\alpha}a^\dagger\ket{\alpha}\bra{\alpha+1},\nonumber\\
\eea
\be\label{eq:commSwithT1}
\left[S,\sum_j\epsilon_{1,j}\ket{j}\bra{j}\right]=\sum_j\gamma_{1,j}(\epsilon_{1,j+1}-\epsilon_{1,j})a^\dagger\ket{j}\bra{j+1},
\ee
\be\label{eq:commSwithT2}
\left[S,\sum_\alpha \epsilon_{2,\alpha}\ket{\alpha}\bra{\alpha}\right]=\sum_\alpha\gamma_{2,\alpha}(\epsilon_{2,\alpha+1}-\epsilon_{2,\alpha})a^\dagger\ket{\alpha}\bra{\alpha+1},
\ee
\bea
&&\left[S,\sum_jh_{1,j}a^\dagger\ket{j}\bra{j+1}\right]\nonumber\\&&=\sum_j(\gamma_{1,j}h_{1,j+1}-\gamma_{1,j+1}h_{1,j})(a^\dagger)^2\ket{j}\bra{j+2},
\eea
\bea
&&\left[S,\sum_jh_{1,j}a\ket{j+1}\bra{j}\right]\nonumber\\&&=\sum_j\gamma_{1,j}h_{1,j}\left(a^\dagger a\ket{j}\bra{j}-aa^\dagger\ket{j+1}\bra{j+1}\right)\nonumber\\
&&-\sum_{j\alpha}\gamma_{2,\alpha}h_{1,j}\ket{j+1}\bra{j}\otimes\ket{\alpha}\bra{\alpha+1},
\eea
\bea
&&\left[S,\sum_\alpha h_{2,\alpha}a^\dagger\ket{\alpha}\bra{\alpha+1}\right]\nonumber\\&&=\sum_\alpha(\gamma_{2,\alpha}h_{2,\alpha+1}-\gamma_{2,\alpha+1}h_{2,\alpha})(a^\dagger)^2\ket{\alpha}\bra{\alpha+2},
\eea
\bea
&&\left[S,\sum_\alpha h_{2,\alpha}a\ket{\alpha+1}\bra{\alpha}\right]\nonumber\\&&=\sum_\alpha\gamma_{2,\alpha}h_{2,\alpha}\left(a^\dagger a\ket{\alpha}\bra{\alpha}-aa^\dagger\ket{\alpha+1}\bra{\alpha+1}\right)\nonumber\\&&-\;\sum_{j\alpha}\gamma_{1,j}h_{2,\alpha}\ket{j}\bra{j+1}\otimes\ket{\alpha+1}\bra{\alpha},
\eea
\begin{widetext}
\bea
\left[S,\left[S-S^\dagger,\omega_c a^\dagger a\right]\right]&=&\left[S,-\omega_c\sum_j\gamma_{1,j}\left(a^\dagger\ket{j}\bra{j+1}+a\ket{j+1}\bra{j}\right)\right]+
\left[S,-\omega_c\sum_\alpha\gamma_{2,\alpha}\left(a^\dagger\ket{\alpha}\bra{\alpha+1}+a\ket{\alpha+1}\bra{\alpha}\right)\right]\nonumber\\
&=&-\omega_c\sum_j\gamma_{1,j}^2\left(a^\dagger a\ket{j}\bra{j}-aa^\dagger\ket{j+1}\bra{j+1}\right)-\omega_c\sum_\alpha\gamma_{2,\alpha}^2\left(a^\dagger a\ket{\alpha}\bra{\alpha}-aa^\dagger\ket{\alpha+1}\bra{\alpha+1}\right)\nonumber\\&&+\;\omega_c\sum_{j\alpha}\gamma_{1,j}\gamma_{2,\alpha}\left(\ket{j+1}\bra{j}\otimes\ket{\alpha}\bra{\alpha+1}+\ket{j}\bra{j+1}\otimes\ket{\alpha+1}\bra{\alpha}\right),\nonumber\\
\eea
\bea
\left[S,\left[S-S^\dagger,\sum_j\epsilon_{1,j}\ket{j}\bra{j}\right]\right]&=&\left[S,\sum_j\gamma_{1,j}(\epsilon_{1,j+1}-\epsilon_{1,j})\left(a^\dagger\ket{j}\bra{j+1}+a\ket{j+1}\bra{j}\right)\right]\nonumber\\&=&\sum_j\gamma_{1,j}\gamma_{1,j+1}(\epsilon_{1,j+2}-2\epsilon_{1,j+1}+\epsilon_{1,j})(a^\dagger)^2\ket{j}\bra{j+2}\nonumber\\
&&+\;\sum_j\gamma_{1,j}^2(\epsilon_{1,j+1}-\epsilon_{1,j})\left(a^\dagger a\ket{j}\bra{j}-aa^\dagger\ket{j+1}\bra{j+1}\right)\nonumber\\
&&-\;\sum_{j\alpha}\gamma_{1,j}\gamma_{2,\alpha}(\epsilon_{1,j+1}-\epsilon_{1,j})\ket{j+1}\bra{j}\otimes\ket{\alpha}\bra{\alpha+1},
\eea
\bea
\left[S,\left[S-S^\dagger,\sum_\alpha\epsilon_{2,\alpha}\ket{\alpha}\bra{\alpha}\right]\right]&=&\left[S,\sum_\alpha\gamma_{2,\alpha}(\epsilon_{2,\alpha+1}-\epsilon_{2,\alpha})\left(a^\dagger\ket{\alpha}\bra{\alpha+1}+a\ket{\alpha+1}\bra{\alpha}\right)\right]\nonumber\\
&=&\sum_\alpha\gamma_{2,\alpha}\gamma_{2,\alpha+1}(\epsilon_{2,\alpha+2}-2\epsilon_{2,\alpha+1}+\epsilon_{2,\alpha})(a^\dagger)^2\ket{\alpha}\bra{\alpha+2}\nonumber\\
    &&+\;\sum_\alpha\gamma_{2,\alpha}^2(\epsilon_{2,\alpha+1}-\epsilon_{2,\alpha})\left(a^\dagger a\ket{\alpha}\bra{\alpha}-aa^\dagger\ket{\alpha+1}\bra{\alpha+1}\right)\nonumber\\
    &&-\;\sum_{j\alpha}\gamma_{1,j}\gamma_{2,\alpha}(\epsilon_{2,\alpha+1}-\epsilon_{2,\alpha})\ket{j}\bra{j+1}\otimes\ket{\alpha+1}\bra{\alpha}.
\eea
\end{widetext}
Eqs.~\eqref{eq:commSwithCavity}-\eqref{eq:commSwithT2} yield the desired values of $\gamma_{1,j}$ and $\gamma_{2,\alpha}$:
\be
\gamma_{1,j}=-\frac{h_{1,j}}{\epsilon_{1,j+1}-\epsilon_{1,j}-\omega_c},\quad \gamma_{2,\alpha}=-\frac{h_{2,\alpha}}{\epsilon_{2,\alpha+1}-\epsilon_{2,\alpha}-\omega_c}.
\ee

If we assume zero cavity photons, then we obtain the following effective Hamiltonian:
\bea\label{eq:hamResonatorAsCapacitive}
H'&\approx&\sum_j\tilde{\epsilon}_{1,j}\ket{j}\bra{j}+\sum_\alpha\tilde{\epsilon}_{2,\alpha}\ket{\alpha}\bra{\alpha}\nonumber\\&&+\;\sum_{j\alpha}\tilde{g}_{j,\alpha}\ket{j}\bra{j+1}\otimes\ket{\alpha+1}\bra{\alpha}+\hbox{H.c.},
\eea
where $\tilde{\epsilon}_{1,0}=\epsilon_{1,0}$, $\tilde{\epsilon}_{2,0}=\epsilon_{2,0}$, and
\bea
\tilde{\epsilon}_{1,j}&=&\epsilon_{1,j}-2\gamma_{1,j-1}h_{1,j-1}+(\omega_c-\epsilon_{1,j}+\epsilon_{1,j-1})\gamma_{1,j-1}^2\nonumber\\&=&\epsilon_{1,j}+\frac{h_{1,j-1}^2}{\epsilon_{1,j}-\epsilon_{1,j-1}-\omega_c},\quad\hbox{for}\quad j>0,
\eea
\bea
\tilde{\epsilon}_{2,\alpha}&=&\epsilon_{2,\alpha}-2\gamma_{2,\alpha-1}h_{2,\alpha-1}+(\omega_c-\epsilon_{2,\alpha}+\epsilon_{2,\alpha-1})\gamma_{2,\alpha-1}^2\nonumber\\&=&\epsilon_{2,\alpha}+\frac{h_{2,\alpha-1}^2}{\epsilon_{2,\alpha}-\epsilon_{2,\alpha-1}-\omega_c},\quad\hbox{for}\quad \alpha>0,
\eea
\bea
\tilde{g}_{j,\alpha}&=&-\gamma_{2,\alpha}h_{1,j}-\gamma_{1,j}h_{2,\alpha}\nonumber\\&&+\;\frac{1}{2}(2\omega_c-\epsilon_{1,j+1}+\epsilon_{1,j}-\epsilon_{2,\alpha+1}+\epsilon_{2,\alpha})\gamma_{1,j}\gamma_{2,\alpha}\nonumber\\
&=&\frac{h_{1,j}h_{2,\alpha}(\epsilon_{1,j+1}-\epsilon_{1,j}+\epsilon_{2,\alpha+1}-\epsilon_{2,\alpha}-2\omega_c)}{2(\epsilon_{1,j+1}-\epsilon_{1,j}-\omega_c)(\epsilon_{2,\alpha+1}-\epsilon_{2,\alpha}-\omega_c)}.
\eea
Eq.~\eqref{eq:hamResonatorAsCapacitive} has the same form as the Hamiltonian describing the direct capacitive coupling between two transmons, Eq.~\eqref{eq:capacitiveHam}. We can therefore use the results from the previous section to write down the effective ZZ-coupling Hamiltonian for two cavity-coupled transmons:
\begin{widetext}
\be
H_{eff}=\left(\begin{matrix}\tilde{\epsilon}_{1,0}+\tilde{\epsilon}_{2,0} & 0 & 0 & 0 \cr 0 & \tilde{\epsilon}_{1,0}+\tilde{\epsilon}_{2,1}+\tilde\eta & 0 & 0 \cr 0 & 0 & \tilde{\epsilon}_{1,1}+\tilde{\epsilon}_{2,0}-\tilde\eta & 0 \cr 0 & 0 & 0 & \tilde{\epsilon}_{1,1}+\tilde{\epsilon}_{2,1}+\tilde{\lambda}_{ZZ} \end{matrix}\right),
\ee
\end{widetext}
where
\be
\tilde\eta=\frac{\tilde{g}_{0,0}^2}{\tilde{\epsilon}_{1,0}+\tilde{\epsilon}_{2,1}-\tilde{\epsilon}_{1,1}-\tilde{\epsilon}_{2,0}},
\ee
\be
\tilde{\lambda}_{ZZ}=\frac{\tilde{g}_{1,0}^2}{\tilde{\epsilon}_{1,1}+\tilde{\epsilon}_{2,1}-\tilde{\epsilon}_{1,2}-\tilde{\epsilon}_{2,0}}+\frac{\tilde{g}_{0,1}^2}{\tilde{\epsilon}_{1,1}+\tilde{\epsilon}_{2,1}-\tilde{\epsilon}_{1,0}-\tilde{\epsilon}_{2,2}}.
\ee
If we define $\omega_1=\tilde{\epsilon}_{1,1}-\tilde{\epsilon}_{1,0}-\tilde{\eta}$, $\omega_2=\tilde{\epsilon}_{2,1}-\tilde{\epsilon}_{2,0}+\tilde{\eta}$, and $g_{z}=\tilde{\lambda}_{ZZ}$, then (after a constant overall energy shift) we obtain Eq.~\ref{2QZZHamiltonian}.

\subsubsection{Driving terms}

We now consider the inclusion of driving terms on each transmon:
\bea
H_d&=&\Omega_1(t)\sum_{j}d_{1,j}\ket{j}\bra{j+1}\nonumber\\&&+\;\Omega_2(t)\sum_{\alpha}d_{2,\alpha}\ket{\alpha}\bra{\alpha+1}+\hbox{H.c.}
\eea
Under the Schrieffer-Wolff transformation, this becomes
\bea
H_d'&=&DH_dD^\dagger=H_d+[S-S^\dagger,H_d]\nonumber\\&&+\;\frac{1}{2}[S-S^\dagger,[S-S^\dagger,H_d]]+\ldots
\eea
The first-order corrections vanish in the zero-cavity-photon subspace because $S$ is proportional to $a^\dagger$, while $H_d$ does not contain any cavity-photon operators. Therefore, the driving terms after the SW transformation are still given by $H_d$.
The full effective Hamiltonian in the logical subspace is
\begin{widetext}
\begin{equation}
\resizebox{1.0 \textwidth}{!}{$
H'_{logical}=\begin{pmatrix} \tilde\epsilon_{1,0}+\tilde\epsilon_{2,0} & d_{2,0}\Omega_2-d_{1,0}\tilde\beta_{0,0}\Omega_1 & d_{1,0}\Omega_1+d_{2,0}\tilde\beta_{0,0}\Omega_2 & 0 \cr d_{2,0}^*\Omega_2^*-d_{1,0}^*\tilde\beta_{0,0}\Omega_1^* & \tilde\epsilon_{1,0}+\tilde\epsilon_{2,1}+\tilde\eta & 0 & d_{1,0}\Omega_1+(d_{2,1}\tilde\beta_{0,1}-d_{2,0}\tilde\beta_{0,0})\Omega_2 \cr d_{1,0}^*\Omega_1^*+d_{2,0}^*\tilde\beta_{0,0}\Omega_2^* & 0 & \tilde\epsilon_{1,1}+\tilde\epsilon_{2,0}-\tilde\eta & (d_{1,0}\tilde\beta_{0,0}-d_{1,1}\tilde\beta_{1,0})\Omega_1+d_{2,0}\Omega_2 \cr 0 & d_{1,0}^*\Omega_1^*+(d_{2,1}^*\tilde\beta_{0,1}-d_{2,0}^*\tilde\beta_{0,0})\Omega_2^* & (d_{1,0}^*\tilde\beta_{0,0}-d_{1,1}^*\tilde\beta_{1,0})\Omega_1^*+d_{2,0}^*\Omega_2^* & \tilde\epsilon_{tr1,1}+\tilde\epsilon_{2,1}+\tilde\lambda_{ZZ} \end{pmatrix},\nonumber\\
$}
\end{equation}
\end{widetext}
where
\be
\tilde\beta_{j,\alpha}=-\frac{\tilde g_{j,\alpha}}{\tilde\epsilon_{1,j}-\tilde\epsilon_{1,j+1}+\tilde\epsilon_{2,\alpha+1}-\tilde\epsilon_{2,\alpha}}.
\ee

\bibliography{SWIPHT}

\begin{thebibliography}{34}%
\makeatletter
\providecommand \@ifxundefined [1]{%
 \@ifx{#1\undefined}
}%
\providecommand \@ifnum [1]{%
 \ifnum #1\expandafter \@firstoftwo
 \else \expandafter \@secondoftwo
 \fi
}%
\providecommand \@ifx [1]{%
 \ifx #1\expandafter \@firstoftwo
 \else \expandafter \@secondoftwo
 \fi
}%
\providecommand \natexlab [1]{#1}%
\providecommand \enquote  [1]{``#1''}%
\providecommand \bibnamefont  [1]{#1}%
\providecommand \bibfnamefont [1]{#1}%
\providecommand \citenamefont [1]{#1}%
\providecommand \href@noop [0]{\@secondoftwo}%
\providecommand \href [0]{\begingroup \@sanitize@url \@href}%
\providecommand \@href[1]{\@@startlink{#1}\@@href}%
\providecommand \@@href[1]{\endgroup#1\@@endlink}%
\providecommand \@sanitize@url [0]{\catcode `\\12\catcode `\$12\catcode
  `\&12\catcode `\#12\catcode `\^12\catcode `\_12\catcode `\%12\relax}%
\providecommand \@@startlink[1]{}%
\providecommand \@@endlink[0]{}%
\providecommand \url  [0]{\begingroup\@sanitize@url \@url }%
\providecommand \@url [1]{\endgroup\@href {#1}{\urlprefix }}%
\providecommand \urlprefix  [0]{URL }%
\providecommand \Eprint [0]{\href }%
\providecommand \doibase [0]{http://dx.doi.org/}%
\providecommand \selectlanguage [0]{\@gobble}%
\providecommand \bibinfo  [0]{\@secondoftwo}%
\providecommand \bibfield  [0]{\@secondoftwo}%
\providecommand \translation [1]{[#1]}%
\providecommand \BibitemOpen [0]{}%
\providecommand \bibitemStop [0]{}%
\providecommand \bibitemNoStop [0]{.\EOS\space}%
\providecommand \EOS [0]{\spacefactor3000\relax}%
\providecommand \BibitemShut  [1]{\csname bibitem#1\endcsname}%
\let\auto@bib@innerbib\@empty
\bibitem [{\citenamefont {Gambetta}\ \emph {et~al.}(2017)\citenamefont
  {Gambetta}, \citenamefont {Chow},\ and\ \citenamefont
  {Steffen}}]{gambetta2017building}%
  \BibitemOpen
  \bibfield  {author} {\bibinfo {author} {\bibfnamefont {Jay~M}\ \bibnamefont
  {Gambetta}}, \bibinfo {author} {\bibfnamefont {Jerry~M}\ \bibnamefont
  {Chow}}, \ and\ \bibinfo {author} {\bibfnamefont {Matthias}\ \bibnamefont
  {Steffen}},\ }\bibfield  {title} {\enquote {\bibinfo {title} {Building
  logical qubits in a superconducting quantum computing system},}\ }\href@noop
  {} {\bibfield  {journal} {\bibinfo  {journal} {npj Quantum Information}\
  }\textbf {\bibinfo {volume} {3}},\ \bibinfo {pages} {1--7} (\bibinfo {year}
  {2017})}\BibitemShut {NoStop}%
\bibitem [{\citenamefont {Krantz}\ \emph {et~al.}(2019)\citenamefont {Krantz},
  \citenamefont {Kjaergaard}, \citenamefont {Yan}, \citenamefont {Orlando},
  \citenamefont {Gustavsson},\ and\ \citenamefont
  {Oliver}}]{krantz2019quantum}%
  \BibitemOpen
  \bibfield  {author} {\bibinfo {author} {\bibfnamefont {Philip}\ \bibnamefont
  {Krantz}}, \bibinfo {author} {\bibfnamefont {Morten}\ \bibnamefont
  {Kjaergaard}}, \bibinfo {author} {\bibfnamefont {Fei}\ \bibnamefont {Yan}},
  \bibinfo {author} {\bibfnamefont {Terry~P}\ \bibnamefont {Orlando}}, \bibinfo
  {author} {\bibfnamefont {Simon}\ \bibnamefont {Gustavsson}}, \ and\ \bibinfo
  {author} {\bibfnamefont {William~D}\ \bibnamefont {Oliver}},\ }\bibfield
  {title} {\enquote {\bibinfo {title} {A quantum engineer's guide to
  superconducting qubits},}\ }\href@noop {} {\bibfield  {journal} {\bibinfo
  {journal} {Applied Physics Reviews}\ }\textbf {\bibinfo {volume} {6}},\
  \bibinfo {pages} {021318} (\bibinfo {year} {2019})}\BibitemShut {NoStop}%
\bibitem [{\citenamefont {Devoret}\ and\ \citenamefont
  {Schoelkopf}(2013)}]{devoret2013superconducting}%
  \BibitemOpen
  \bibfield  {author} {\bibinfo {author} {\bibfnamefont {Michel~H}\
  \bibnamefont {Devoret}}\ and\ \bibinfo {author} {\bibfnamefont {Robert~J}\
  \bibnamefont {Schoelkopf}},\ }\bibfield  {title} {\enquote {\bibinfo {title}
  {Superconducting circuits for quantum information: an outlook},}\ }\href@noop
  {} {\bibfield  {journal} {\bibinfo  {journal} {Science}\ }\textbf {\bibinfo
  {volume} {339}},\ \bibinfo {pages} {1169--1174} (\bibinfo {year}
  {2013})}\BibitemShut {NoStop}%
\bibitem [{\citenamefont {Kjaergaard}\ \emph {et~al.}(2019)\citenamefont
  {Kjaergaard}, \citenamefont {Schwartz}, \citenamefont {Braum{\"u}ller},
  \citenamefont {Krantz}, \citenamefont {Wang}, \citenamefont {Gustavsson},\
  and\ \citenamefont {Oliver}}]{kjaergaard2019superconducting}%
  \BibitemOpen
  \bibfield  {author} {\bibinfo {author} {\bibfnamefont {Morten}\ \bibnamefont
  {Kjaergaard}}, \bibinfo {author} {\bibfnamefont {Mollie~E}\ \bibnamefont
  {Schwartz}}, \bibinfo {author} {\bibfnamefont {Jochen}\ \bibnamefont
  {Braum{\"u}ller}}, \bibinfo {author} {\bibfnamefont {Philip}\ \bibnamefont
  {Krantz}}, \bibinfo {author} {\bibfnamefont {Joel I-J}\ \bibnamefont {Wang}},
  \bibinfo {author} {\bibfnamefont {Simon}\ \bibnamefont {Gustavsson}}, \ and\
  \bibinfo {author} {\bibfnamefont {William~D}\ \bibnamefont {Oliver}},\
  }\bibfield  {title} {\enquote {\bibinfo {title} {Superconducting qubits:
  Current state of play},}\ }\href@noop {} {\bibfield  {journal} {\bibinfo
  {journal} {Annual Review of Condensed Matter Physics}\ }\textbf {\bibinfo
  {volume} {11}} (\bibinfo {year} {2019})}\BibitemShut {NoStop}%
\bibitem [{\citenamefont {Barends}\ \emph {et~al.}(2014)\citenamefont
  {Barends}, \citenamefont {Kelly}, \citenamefont {Megrant}, \citenamefont
  {Veitia}, \citenamefont {Sank}, \citenamefont {Jeffrey}, \citenamefont
  {White}, \citenamefont {Mutus}, \citenamefont {Fowler}, \citenamefont
  {Campbell} \emph {et~al.}}]{barends2014superconducting}%
  \BibitemOpen
  \bibfield  {author} {\bibinfo {author} {\bibfnamefont {Rami}\ \bibnamefont
  {Barends}}, \bibinfo {author} {\bibfnamefont {Julian}\ \bibnamefont {Kelly}},
  \bibinfo {author} {\bibfnamefont {Anthony}\ \bibnamefont {Megrant}}, \bibinfo
  {author} {\bibfnamefont {Andrzej}\ \bibnamefont {Veitia}}, \bibinfo {author}
  {\bibfnamefont {Daniel}\ \bibnamefont {Sank}}, \bibinfo {author}
  {\bibfnamefont {Evan}\ \bibnamefont {Jeffrey}}, \bibinfo {author}
  {\bibfnamefont {Ted~C}\ \bibnamefont {White}}, \bibinfo {author}
  {\bibfnamefont {Josh}\ \bibnamefont {Mutus}}, \bibinfo {author}
  {\bibfnamefont {Austin~G}\ \bibnamefont {Fowler}}, \bibinfo {author}
  {\bibfnamefont {Brooks}\ \bibnamefont {Campbell}},  \emph {et~al.},\
  }\bibfield  {title} {\enquote {\bibinfo {title} {Superconducting quantum
  circuits at the surface code threshold for fault tolerance},}\ }\href@noop {}
  {\bibfield  {journal} {\bibinfo  {journal} {Nature}\ }\textbf {\bibinfo
  {volume} {508}},\ \bibinfo {pages} {500--503} (\bibinfo {year}
  {2014})}\BibitemShut {NoStop}%
\bibitem [{\citenamefont {McKay}\ \emph {et~al.}(2019)\citenamefont {McKay},
  \citenamefont {Sheldon}, \citenamefont {Smolin}, \citenamefont {Chow},\ and\
  \citenamefont {Gambetta}}]{mckay2019three}%
  \BibitemOpen
  \bibfield  {author} {\bibinfo {author} {\bibfnamefont {David~C}\ \bibnamefont
  {McKay}}, \bibinfo {author} {\bibfnamefont {Sarah}\ \bibnamefont {Sheldon}},
  \bibinfo {author} {\bibfnamefont {John~A}\ \bibnamefont {Smolin}}, \bibinfo
  {author} {\bibfnamefont {Jerry~M}\ \bibnamefont {Chow}}, \ and\ \bibinfo
  {author} {\bibfnamefont {Jay~M}\ \bibnamefont {Gambetta}},\ }\bibfield
  {title} {\enquote {\bibinfo {title} {Three-qubit randomized benchmarking},}\
  }\href@noop {} {\bibfield  {journal} {\bibinfo  {journal} {Physical review
  letters}\ }\textbf {\bibinfo {volume} {122}},\ \bibinfo {pages} {200502}
  (\bibinfo {year} {2019})}\BibitemShut {NoStop}%
\bibitem [{\citenamefont {Hong}\ \emph {et~al.}(2020)\citenamefont {Hong},
  \citenamefont {Papageorge}, \citenamefont {Sivarajah}, \citenamefont
  {Crossman}, \citenamefont {Didier}, \citenamefont {Polloreno}, \citenamefont
  {Sete}, \citenamefont {Turkowski}, \citenamefont {da~Silva},\ and\
  \citenamefont {Johnson}}]{hong2020demonstration}%
  \BibitemOpen
  \bibfield  {author} {\bibinfo {author} {\bibfnamefont {Sabrina~S}\
  \bibnamefont {Hong}}, \bibinfo {author} {\bibfnamefont {Alexander~T}\
  \bibnamefont {Papageorge}}, \bibinfo {author} {\bibfnamefont {Prasahnt}\
  \bibnamefont {Sivarajah}}, \bibinfo {author} {\bibfnamefont {Genya}\
  \bibnamefont {Crossman}}, \bibinfo {author} {\bibfnamefont {Nicolas}\
  \bibnamefont {Didier}}, \bibinfo {author} {\bibfnamefont {Anthony~M}\
  \bibnamefont {Polloreno}}, \bibinfo {author} {\bibfnamefont {Eyob~A}\
  \bibnamefont {Sete}}, \bibinfo {author} {\bibfnamefont {Stefan~W}\
  \bibnamefont {Turkowski}}, \bibinfo {author} {\bibfnamefont {Marcus~P}\
  \bibnamefont {da~Silva}}, \ and\ \bibinfo {author} {\bibfnamefont {Blake~R}\
  \bibnamefont {Johnson}},\ }\bibfield  {title} {\enquote {\bibinfo {title}
  {Demonstration of a parametrically activated entangling gate protected from
  flux noise},}\ }\href@noop {} {\bibfield  {journal} {\bibinfo  {journal}
  {Physical Review A}\ }\textbf {\bibinfo {volume} {101}},\ \bibinfo {pages}
  {012302} (\bibinfo {year} {2020})}\BibitemShut {NoStop}%
\bibitem [{\citenamefont {Cross}\ \emph {et~al.}(2019)\citenamefont {Cross},
  \citenamefont {Bishop}, \citenamefont {Sheldon}, \citenamefont {Nation},\
  and\ \citenamefont {Gambetta}}]{cross2019validating}%
  \BibitemOpen
  \bibfield  {author} {\bibinfo {author} {\bibfnamefont {Andrew~W}\
  \bibnamefont {Cross}}, \bibinfo {author} {\bibfnamefont {Lev~S}\ \bibnamefont
  {Bishop}}, \bibinfo {author} {\bibfnamefont {Sarah}\ \bibnamefont {Sheldon}},
  \bibinfo {author} {\bibfnamefont {Paul~D}\ \bibnamefont {Nation}}, \ and\
  \bibinfo {author} {\bibfnamefont {Jay~M}\ \bibnamefont {Gambetta}},\
  }\bibfield  {title} {\enquote {\bibinfo {title} {Validating quantum computers
  using randomized model circuits},}\ }\href@noop {} {\bibfield  {journal}
  {\bibinfo  {journal} {Physical Review A}\ }\textbf {\bibinfo {volume}
  {100}},\ \bibinfo {pages} {032328} (\bibinfo {year} {2019})}\BibitemShut
  {NoStop}%
\bibitem [{\citenamefont {Arute}\ \emph {et~al.}(2019)\citenamefont {Arute},
  \citenamefont {Arya}, \citenamefont {Babbush}, \citenamefont {Bacon},
  \citenamefont {Bardin}, \citenamefont {Barends}, \citenamefont {Biswas},
  \citenamefont {Boixo}, \citenamefont {Brandao}, \citenamefont {Buell} \emph
  {et~al.}}]{arute2019quantum}%
  \BibitemOpen
  \bibfield  {author} {\bibinfo {author} {\bibfnamefont {Frank}\ \bibnamefont
  {Arute}}, \bibinfo {author} {\bibfnamefont {Kunal}\ \bibnamefont {Arya}},
  \bibinfo {author} {\bibfnamefont {Ryan}\ \bibnamefont {Babbush}}, \bibinfo
  {author} {\bibfnamefont {Dave}\ \bibnamefont {Bacon}}, \bibinfo {author}
  {\bibfnamefont {Joseph~C}\ \bibnamefont {Bardin}}, \bibinfo {author}
  {\bibfnamefont {Rami}\ \bibnamefont {Barends}}, \bibinfo {author}
  {\bibfnamefont {Rupak}\ \bibnamefont {Biswas}}, \bibinfo {author}
  {\bibfnamefont {Sergio}\ \bibnamefont {Boixo}}, \bibinfo {author}
  {\bibfnamefont {Fernando~GSL}\ \bibnamefont {Brandao}}, \bibinfo {author}
  {\bibfnamefont {David~A}\ \bibnamefont {Buell}},  \emph {et~al.},\ }\bibfield
   {title} {\enquote {\bibinfo {title} {Quantum supremacy using a programmable
  superconducting processor},}\ }\href@noop {} {\bibfield  {journal} {\bibinfo
  {journal} {Nature}\ }\textbf {\bibinfo {volume} {574}},\ \bibinfo {pages}
  {505--510} (\bibinfo {year} {2019})}\BibitemShut {NoStop}%
\bibitem [{\citenamefont {Otterbach}\ \emph {et~al.}(2017)\citenamefont
  {Otterbach}, \citenamefont {Manenti}, \citenamefont {Alidoust}, \citenamefont
  {Bestwick}, \citenamefont {Block}, \citenamefont {Bloom}, \citenamefont
  {Caldwell}, \citenamefont {Didier}, \citenamefont {Fried}, \citenamefont
  {Hong} \emph {et~al.}}]{otterbach2017unsupervised}%
  \BibitemOpen
  \bibfield  {author} {\bibinfo {author} {\bibfnamefont {JS}~\bibnamefont
  {Otterbach}}, \bibinfo {author} {\bibfnamefont {R}~\bibnamefont {Manenti}},
  \bibinfo {author} {\bibfnamefont {N}~\bibnamefont {Alidoust}}, \bibinfo
  {author} {\bibfnamefont {A}~\bibnamefont {Bestwick}}, \bibinfo {author}
  {\bibfnamefont {M}~\bibnamefont {Block}}, \bibinfo {author} {\bibfnamefont
  {B}~\bibnamefont {Bloom}}, \bibinfo {author} {\bibfnamefont {S}~\bibnamefont
  {Caldwell}}, \bibinfo {author} {\bibfnamefont {N}~\bibnamefont {Didier}},
  \bibinfo {author} {\bibfnamefont {E~Schuyler}\ \bibnamefont {Fried}},
  \bibinfo {author} {\bibfnamefont {S}~\bibnamefont {Hong}},  \emph {et~al.},\
  }\bibfield  {title} {\enquote {\bibinfo {title} {Unsupervised machine
  learning on a hybrid quantum computer},}\ }\href@noop {} {\bibfield
  {journal} {\bibinfo  {journal} {arXiv preprint arXiv:1712.05771}\ } (\bibinfo
  {year} {2017})}\BibitemShut {NoStop}%
\bibitem [{\citenamefont {Motzoi}\ \emph {et~al.}(2009)\citenamefont {Motzoi},
  \citenamefont {Gambetta}, \citenamefont {Rebentrost},\ and\ \citenamefont
  {Wilhelm}}]{motzoi2009simple}%
  \BibitemOpen
  \bibfield  {author} {\bibinfo {author} {\bibfnamefont {Felix}\ \bibnamefont
  {Motzoi}}, \bibinfo {author} {\bibfnamefont {Jay~M}\ \bibnamefont
  {Gambetta}}, \bibinfo {author} {\bibfnamefont {Patrick}\ \bibnamefont
  {Rebentrost}}, \ and\ \bibinfo {author} {\bibfnamefont {Frank~K}\
  \bibnamefont {Wilhelm}},\ }\bibfield  {title} {\enquote {\bibinfo {title}
  {Simple pulses for elimination of leakage in weakly nonlinear qubits},}\
  }\href@noop {} {\bibfield  {journal} {\bibinfo  {journal} {Physical review
  letters}\ }\textbf {\bibinfo {volume} {103}},\ \bibinfo {pages} {110501}
  (\bibinfo {year} {2009})}\BibitemShut {NoStop}%
\bibitem [{\citenamefont {Sheldon}\ \emph {et~al.}(2016)\citenamefont
  {Sheldon}, \citenamefont {Bishop}, \citenamefont {Magesan}, \citenamefont
  {Filipp}, \citenamefont {Chow},\ and\ \citenamefont
  {Gambetta}}]{sheldon2016characterizing}%
  \BibitemOpen
  \bibfield  {author} {\bibinfo {author} {\bibfnamefont {Sarah}\ \bibnamefont
  {Sheldon}}, \bibinfo {author} {\bibfnamefont {Lev~S}\ \bibnamefont {Bishop}},
  \bibinfo {author} {\bibfnamefont {Easwar}\ \bibnamefont {Magesan}}, \bibinfo
  {author} {\bibfnamefont {Stefan}\ \bibnamefont {Filipp}}, \bibinfo {author}
  {\bibfnamefont {Jerry~M}\ \bibnamefont {Chow}}, \ and\ \bibinfo {author}
  {\bibfnamefont {Jay~M}\ \bibnamefont {Gambetta}},\ }\bibfield  {title}
  {\enquote {\bibinfo {title} {Characterizing errors on qubit operations via
  iterative randomized benchmarking},}\ }\href@noop {} {\bibfield  {journal}
  {\bibinfo  {journal} {Physical Review A}\ }\textbf {\bibinfo {volume} {93}},\
  \bibinfo {pages} {012301} (\bibinfo {year} {2016})}\BibitemShut {NoStop}%
\bibitem [{\citenamefont {Kelly}\ \emph {et~al.}(2014)\citenamefont {Kelly},
  \citenamefont {Barends}, \citenamefont {Campbell}, \citenamefont {Chen},
  \citenamefont {Chen}, \citenamefont {Chiaro}, \citenamefont {Dunsworth},
  \citenamefont {Fowler}, \citenamefont {Hoi}, \citenamefont {Jeffrey} \emph
  {et~al.}}]{kelly2014optimal}%
  \BibitemOpen
  \bibfield  {author} {\bibinfo {author} {\bibfnamefont {Julian}\ \bibnamefont
  {Kelly}}, \bibinfo {author} {\bibfnamefont {R}~\bibnamefont {Barends}},
  \bibinfo {author} {\bibfnamefont {B}~\bibnamefont {Campbell}}, \bibinfo
  {author} {\bibfnamefont {Y}~\bibnamefont {Chen}}, \bibinfo {author}
  {\bibfnamefont {Z}~\bibnamefont {Chen}}, \bibinfo {author} {\bibfnamefont
  {B}~\bibnamefont {Chiaro}}, \bibinfo {author} {\bibfnamefont {A}~\bibnamefont
  {Dunsworth}}, \bibinfo {author} {\bibfnamefont {Austin~G}\ \bibnamefont
  {Fowler}}, \bibinfo {author} {\bibfnamefont {I-C}\ \bibnamefont {Hoi}},
  \bibinfo {author} {\bibfnamefont {E}~\bibnamefont {Jeffrey}},  \emph
  {et~al.},\ }\bibfield  {title} {\enquote {\bibinfo {title} {Optimal quantum
  control using randomized benchmarking},}\ }\href@noop {} {\bibfield
  {journal} {\bibinfo  {journal} {Physical review letters}\ }\textbf {\bibinfo
  {volume} {112}},\ \bibinfo {pages} {240504} (\bibinfo {year}
  {2014})}\BibitemShut {NoStop}%
\bibitem [{\citenamefont {Foxen}\ \emph {et~al.}(2020)\citenamefont {Foxen},
  \citenamefont {Neill}, \citenamefont {Dunsworth}, \citenamefont {Roushan},
  \citenamefont {Chiaro}, \citenamefont {Megrant}, \citenamefont {Kelly},
  \citenamefont {Chen}, \citenamefont {Satzinger}, \citenamefont {Barends}
  \emph {et~al.}}]{foxen2020demonstrating}%
  \BibitemOpen
  \bibfield  {author} {\bibinfo {author} {\bibfnamefont {Brooks}\ \bibnamefont
  {Foxen}}, \bibinfo {author} {\bibfnamefont {Charles}\ \bibnamefont {Neill}},
  \bibinfo {author} {\bibfnamefont {Andrew}\ \bibnamefont {Dunsworth}},
  \bibinfo {author} {\bibfnamefont {Pedram}\ \bibnamefont {Roushan}}, \bibinfo
  {author} {\bibfnamefont {Ben}\ \bibnamefont {Chiaro}}, \bibinfo {author}
  {\bibfnamefont {Anthony}\ \bibnamefont {Megrant}}, \bibinfo {author}
  {\bibfnamefont {Julian}\ \bibnamefont {Kelly}}, \bibinfo {author}
  {\bibfnamefont {Zijun}\ \bibnamefont {Chen}}, \bibinfo {author}
  {\bibfnamefont {Kevin}\ \bibnamefont {Satzinger}}, \bibinfo {author}
  {\bibfnamefont {Rami}\ \bibnamefont {Barends}},  \emph {et~al.},\ }\bibfield
  {title} {\enquote {\bibinfo {title} {Demonstrating a continuous set of
  two-qubit gates for near-term quantum algorithms},}\ }\href@noop {}
  {\bibfield  {journal} {\bibinfo  {journal} {arXiv preprint arXiv:2001.08343}\
  } (\bibinfo {year} {2020})}\BibitemShut {NoStop}%
\bibitem [{\citenamefont {Kjaergaard}\ \emph {et~al.}(2020)\citenamefont
  {Kjaergaard}, \citenamefont {Schwartz}, \citenamefont {Greene}, \citenamefont
  {Samach}, \citenamefont {Bengtsson}, \citenamefont {O'Keeffe}, \citenamefont
  {McNally}, \citenamefont {Braum{\"u}ller}, \citenamefont {Kim}, \citenamefont
  {Krantz} \emph {et~al.}}]{kjaergaard2020quantum}%
  \BibitemOpen
  \bibfield  {author} {\bibinfo {author} {\bibfnamefont {M}~\bibnamefont
  {Kjaergaard}}, \bibinfo {author} {\bibfnamefont {ME}~\bibnamefont
  {Schwartz}}, \bibinfo {author} {\bibfnamefont {A}~\bibnamefont {Greene}},
  \bibinfo {author} {\bibfnamefont {GO}~\bibnamefont {Samach}}, \bibinfo
  {author} {\bibfnamefont {A}~\bibnamefont {Bengtsson}}, \bibinfo {author}
  {\bibfnamefont {M}~\bibnamefont {O'Keeffe}}, \bibinfo {author} {\bibfnamefont
  {CM}~\bibnamefont {McNally}}, \bibinfo {author} {\bibfnamefont
  {J}~\bibnamefont {Braum{\"u}ller}}, \bibinfo {author} {\bibfnamefont
  {DK}~\bibnamefont {Kim}}, \bibinfo {author} {\bibfnamefont {P}~\bibnamefont
  {Krantz}},  \emph {et~al.},\ }\bibfield  {title} {\enquote {\bibinfo {title}
  {A quantum instruction set implemented on a superconducting quantum
  processor},}\ }\href@noop {} {\bibfield  {journal} {\bibinfo  {journal}
  {arXiv preprint arXiv:2001.08838}\ } (\bibinfo {year} {2020})}\BibitemShut
  {NoStop}%
\bibitem [{\citenamefont {Krinner}\ \emph {et~al.}(2020)\citenamefont
  {Krinner}, \citenamefont {Kurpiers}, \citenamefont {Royer}, \citenamefont
  {Magnard}, \citenamefont {Tsitsilin}, \citenamefont {Besse}, \citenamefont
  {Remm}, \citenamefont {Blais},\ and\ \citenamefont
  {Wallraff}}]{krinner2020demonstration}%
  \BibitemOpen
  \bibfield  {author} {\bibinfo {author} {\bibfnamefont {S}~\bibnamefont
  {Krinner}}, \bibinfo {author} {\bibfnamefont {P}~\bibnamefont {Kurpiers}},
  \bibinfo {author} {\bibfnamefont {B}~\bibnamefont {Royer}}, \bibinfo {author}
  {\bibfnamefont {P}~\bibnamefont {Magnard}}, \bibinfo {author} {\bibfnamefont
  {I}~\bibnamefont {Tsitsilin}}, \bibinfo {author} {\bibfnamefont {J-C}\
  \bibnamefont {Besse}}, \bibinfo {author} {\bibfnamefont {A}~\bibnamefont
  {Remm}}, \bibinfo {author} {\bibfnamefont {A}~\bibnamefont {Blais}}, \ and\
  \bibinfo {author} {\bibfnamefont {A}~\bibnamefont {Wallraff}},\ }\bibfield
  {title} {\enquote {\bibinfo {title} {Demonstration of an all-microwave
  controlled-phase gate between far detuned qubits},}\ }\href@noop {}
  {\bibfield  {journal} {\bibinfo  {journal} {arXiv preprint arXiv:2006.10639}\
  } (\bibinfo {year} {2020})}\BibitemShut {NoStop}%
\bibitem [{\citenamefont {Xu}\ \emph {et~al.}(2020)\citenamefont {Xu},
  \citenamefont {Chu}, \citenamefont {Yuan}, \citenamefont {Qiu}, \citenamefont
  {Zhou}, \citenamefont {Zhang}, \citenamefont {Tan}, \citenamefont {Yu},
  \citenamefont {Liu}, \citenamefont {Li} \emph {et~al.}}]{xu2020high}%
  \BibitemOpen
  \bibfield  {author} {\bibinfo {author} {\bibfnamefont {Yuan}\ \bibnamefont
  {Xu}}, \bibinfo {author} {\bibfnamefont {Ji}~\bibnamefont {Chu}}, \bibinfo
  {author} {\bibfnamefont {Jiahao}\ \bibnamefont {Yuan}}, \bibinfo {author}
  {\bibfnamefont {Jiawei}\ \bibnamefont {Qiu}}, \bibinfo {author}
  {\bibfnamefont {Yuxuan}\ \bibnamefont {Zhou}}, \bibinfo {author}
  {\bibfnamefont {Libo}\ \bibnamefont {Zhang}}, \bibinfo {author}
  {\bibfnamefont {Xinsheng}\ \bibnamefont {Tan}}, \bibinfo {author}
  {\bibfnamefont {Yang}\ \bibnamefont {Yu}}, \bibinfo {author} {\bibfnamefont
  {Song}\ \bibnamefont {Liu}}, \bibinfo {author} {\bibfnamefont {Jian}\
  \bibnamefont {Li}},  \emph {et~al.},\ }\bibfield  {title} {\enquote {\bibinfo
  {title} {High-fidelity, high-scalability two-qubit gate scheme for
  superconducting qubits},}\ }\href@noop {} {\bibfield  {journal} {\bibinfo
  {journal} {arXiv preprint arXiv:2006.11860}\ } (\bibinfo {year}
  {2020})}\BibitemShut {NoStop}%
\bibitem [{\citenamefont {Collodo}\ \emph {et~al.}(2020)\citenamefont
  {Collodo}, \citenamefont {Herrmann}, \citenamefont {Lacroix}, \citenamefont
  {Andersen}, \citenamefont {Remm}, \citenamefont {Lazar}, \citenamefont
  {Besse}, \citenamefont {Walter}, \citenamefont {Wallraff},\ and\
  \citenamefont {Eichler}}]{collodo2020implementation}%
  \BibitemOpen
  \bibfield  {author} {\bibinfo {author} {\bibfnamefont {Michele~C}\
  \bibnamefont {Collodo}}, \bibinfo {author} {\bibfnamefont {Johannes}\
  \bibnamefont {Herrmann}}, \bibinfo {author} {\bibfnamefont {Nathan}\
  \bibnamefont {Lacroix}}, \bibinfo {author} {\bibfnamefont
  {Christian~Kraglund}\ \bibnamefont {Andersen}}, \bibinfo {author}
  {\bibfnamefont {Ants}\ \bibnamefont {Remm}}, \bibinfo {author} {\bibfnamefont
  {Stefania}\ \bibnamefont {Lazar}}, \bibinfo {author} {\bibfnamefont
  {Jean-Claude}\ \bibnamefont {Besse}}, \bibinfo {author} {\bibfnamefont
  {Theo}\ \bibnamefont {Walter}}, \bibinfo {author} {\bibfnamefont {Andreas}\
  \bibnamefont {Wallraff}}, \ and\ \bibinfo {author} {\bibfnamefont
  {Christopher}\ \bibnamefont {Eichler}},\ }\bibfield  {title} {\enquote
  {\bibinfo {title} {Implementation of conditional-phase gates based on tunable
  zz-interactions},}\ }\href@noop {} {\bibfield  {journal} {\bibinfo  {journal}
  {arXiv preprint arXiv:2005.08863}\ } (\bibinfo {year} {2020})}\BibitemShut
  {NoStop}%
\bibitem [{\citenamefont {McKay}\ \emph {et~al.}(2016)\citenamefont {McKay},
  \citenamefont {Filipp}, \citenamefont {Mezzacapo}, \citenamefont {Magesan},
  \citenamefont {Chow},\ and\ \citenamefont {Gambetta}}]{mckay2016universal}%
  \BibitemOpen
  \bibfield  {author} {\bibinfo {author} {\bibfnamefont {David~C}\ \bibnamefont
  {McKay}}, \bibinfo {author} {\bibfnamefont {Stefan}\ \bibnamefont {Filipp}},
  \bibinfo {author} {\bibfnamefont {Antonio}\ \bibnamefont {Mezzacapo}},
  \bibinfo {author} {\bibfnamefont {Easwar}\ \bibnamefont {Magesan}}, \bibinfo
  {author} {\bibfnamefont {Jerry~M}\ \bibnamefont {Chow}}, \ and\ \bibinfo
  {author} {\bibfnamefont {Jay~M}\ \bibnamefont {Gambetta}},\ }\bibfield
  {title} {\enquote {\bibinfo {title} {Universal gate for fixed-frequency
  qubits via a tunable bus},}\ }\href@noop {} {\bibfield  {journal} {\bibinfo
  {journal} {Physical Review Applied}\ }\textbf {\bibinfo {volume} {6}},\
  \bibinfo {pages} {064007} (\bibinfo {year} {2016})}\BibitemShut {NoStop}%
\bibitem [{\citenamefont {Mundada}\ \emph {et~al.}(2019)\citenamefont
  {Mundada}, \citenamefont {Zhang}, \citenamefont {Hazard},\ and\ \citenamefont
  {Houck}}]{mundada2019suppression}%
  \BibitemOpen
  \bibfield  {author} {\bibinfo {author} {\bibfnamefont {Pranav}\ \bibnamefont
  {Mundada}}, \bibinfo {author} {\bibfnamefont {Gengyan}\ \bibnamefont
  {Zhang}}, \bibinfo {author} {\bibfnamefont {Thomas}\ \bibnamefont {Hazard}},
  \ and\ \bibinfo {author} {\bibfnamefont {Andrew}\ \bibnamefont {Houck}},\
  }\bibfield  {title} {\enquote {\bibinfo {title} {Suppression of qubit
  crosstalk in a tunable coupling superconducting circuit},}\ }\href@noop {}
  {\bibfield  {journal} {\bibinfo  {journal} {Physical Review Applied}\
  }\textbf {\bibinfo {volume} {12}},\ \bibinfo {pages} {054023} (\bibinfo
  {year} {2019})}\BibitemShut {NoStop}%
\bibitem [{\citenamefont {Li}\ \emph {et~al.}(2020)\citenamefont {Li},
  \citenamefont {Cai}, \citenamefont {Yan}, \citenamefont {Wang}, \citenamefont
  {Pan}, \citenamefont {Ma}, \citenamefont {Cai}, \citenamefont {Han},
  \citenamefont {Hua}, \citenamefont {Han}, \citenamefont {Wu}, \citenamefont
  {Zhang}, \citenamefont {Wang}, \citenamefont {Song}, \citenamefont {Duan},\
  and\ \citenamefont {Sun}}]{Li2020_tunable_coupler}%
  \BibitemOpen
  \bibfield  {author} {\bibinfo {author} {\bibfnamefont {X.}~\bibnamefont
  {Li}}, \bibinfo {author} {\bibfnamefont {T.}~\bibnamefont {Cai}}, \bibinfo
  {author} {\bibfnamefont {H.}~\bibnamefont {Yan}}, \bibinfo {author}
  {\bibfnamefont {Z.}~\bibnamefont {Wang}}, \bibinfo {author} {\bibfnamefont
  {X.}~\bibnamefont {Pan}}, \bibinfo {author} {\bibfnamefont {Y.}~\bibnamefont
  {Ma}}, \bibinfo {author} {\bibfnamefont {W.}~\bibnamefont {Cai}}, \bibinfo
  {author} {\bibfnamefont {J.}~\bibnamefont {Han}}, \bibinfo {author}
  {\bibfnamefont {Z.}~\bibnamefont {Hua}}, \bibinfo {author} {\bibfnamefont
  {X.}~\bibnamefont {Han}}, \bibinfo {author} {\bibfnamefont {Y.}~\bibnamefont
  {Wu}}, \bibinfo {author} {\bibfnamefont {H.}~\bibnamefont {Zhang}}, \bibinfo
  {author} {\bibfnamefont {H.}~\bibnamefont {Wang}}, \bibinfo {author}
  {\bibfnamefont {Yipu}\ \bibnamefont {Song}}, \bibinfo {author} {\bibfnamefont
  {Luming}\ \bibnamefont {Duan}}, \ and\ \bibinfo {author} {\bibfnamefont
  {Luyan}\ \bibnamefont {Sun}},\ }\bibfield  {title} {\enquote {\bibinfo
  {title} {Tunable coupler for realizing a controlled-phase gate with
  dynamically decoupled regime in a superconducting circuit},}\ }\href
  {\doibase 10.1103/PhysRevApplied.14.024070} {\bibfield  {journal} {\bibinfo
  {journal} {Phys. Rev. Applied}\ }\textbf {\bibinfo {volume} {14}},\ \bibinfo
  {pages} {024070} (\bibinfo {year} {2020})}\BibitemShut {NoStop}%
\bibitem [{\citenamefont {Ku}\ \emph {et~al.}(2020)\citenamefont {Ku},
  \citenamefont {Xu}, \citenamefont {Brink}, \citenamefont {McKay},
  \citenamefont {Hertzberg}, \citenamefont {Ansari},\ and\ \citenamefont
  {Plourde}}]{ku2020suppression}%
  \BibitemOpen
  \bibfield  {author} {\bibinfo {author} {\bibfnamefont {Jaseung}\ \bibnamefont
  {Ku}}, \bibinfo {author} {\bibfnamefont {Xuexin}\ \bibnamefont {Xu}},
  \bibinfo {author} {\bibfnamefont {Markus}\ \bibnamefont {Brink}}, \bibinfo
  {author} {\bibfnamefont {David~C}\ \bibnamefont {McKay}}, \bibinfo {author}
  {\bibfnamefont {Jared~B}\ \bibnamefont {Hertzberg}}, \bibinfo {author}
  {\bibfnamefont {Mohammad~H}\ \bibnamefont {Ansari}}, \ and\ \bibinfo {author}
  {\bibfnamefont {BLT}\ \bibnamefont {Plourde}},\ }\bibfield  {title} {\enquote
  {\bibinfo {title} {Suppression of unwanted $ zz $ interactions in a hybrid
  two-qubit system},}\ }\href@noop {} {\bibfield  {journal} {\bibinfo
  {journal} {arXiv preprint arXiv:2003.02775}\ } (\bibinfo {year}
  {2020})}\BibitemShut {NoStop}%
\bibitem [{\citenamefont {Economou}\ and\ \citenamefont
  {Barnes}(2015)}]{economou2015analytical}%
  \BibitemOpen
  \bibfield  {author} {\bibinfo {author} {\bibfnamefont {Sophia~E}\
  \bibnamefont {Economou}}\ and\ \bibinfo {author} {\bibfnamefont {Edwin}\
  \bibnamefont {Barnes}},\ }\bibfield  {title} {\enquote {\bibinfo {title}
  {Analytical approach to swift nonleaky entangling gates in superconducting
  qubits},}\ }\href@noop {} {\bibfield  {journal} {\bibinfo  {journal}
  {Physical Review B}\ }\textbf {\bibinfo {volume} {91}},\ \bibinfo {pages}
  {161405} (\bibinfo {year} {2015})}\BibitemShut {NoStop}%
\bibitem [{\citenamefont {Deng}\ \emph {et~al.}(2017)\citenamefont {Deng},
  \citenamefont {Barnes},\ and\ \citenamefont {Economou}}]{deng2017robustness}%
  \BibitemOpen
  \bibfield  {author} {\bibinfo {author} {\bibfnamefont {Xiu-Hao}\ \bibnamefont
  {Deng}}, \bibinfo {author} {\bibfnamefont {Edwin}\ \bibnamefont {Barnes}}, \
  and\ \bibinfo {author} {\bibfnamefont {Sophia~E}\ \bibnamefont {Economou}},\
  }\bibfield  {title} {\enquote {\bibinfo {title} {Robustness of
  error-suppressing entangling gates in cavity-coupled transmon qubits},}\
  }\href@noop {} {\bibfield  {journal} {\bibinfo  {journal} {Physical Review
  B}\ }\textbf {\bibinfo {volume} {96}},\ \bibinfo {pages} {035441} (\bibinfo
  {year} {2017})}\BibitemShut {NoStop}%
\bibitem [{\citenamefont {Barron}\ \emph {et~al.}(2020)\citenamefont {Barron},
  \citenamefont {Calderon-Vargas}, \citenamefont {Long}, \citenamefont
  {Pappas},\ and\ \citenamefont {Economou}}]{barron2020microwave}%
  \BibitemOpen
  \bibfield  {author} {\bibinfo {author} {\bibfnamefont {George~S.}\
  \bibnamefont {Barron}}, \bibinfo {author} {\bibfnamefont {F.~A.}\
  \bibnamefont {Calderon-Vargas}}, \bibinfo {author} {\bibfnamefont {Junling}\
  \bibnamefont {Long}}, \bibinfo {author} {\bibfnamefont {David~P.}\
  \bibnamefont {Pappas}}, \ and\ \bibinfo {author} {\bibfnamefont {Sophia~E.}\
  \bibnamefont {Economou}},\ }\bibfield  {title} {\enquote {\bibinfo {title}
  {Microwave-based arbitrary cphase gates for transmon qubits},}\ }\href
  {\doibase 10.1103/PhysRevB.101.054508} {\bibfield  {journal} {\bibinfo
  {journal} {Phys. Rev. B}\ }\textbf {\bibinfo {volume} {101}},\ \bibinfo
  {pages} {054508} (\bibinfo {year} {2020})}\BibitemShut {NoStop}%
\bibitem [{\citenamefont {Koch}\ \emph {et~al.}(2007)\citenamefont {Koch},
  \citenamefont {Yu}, \citenamefont {Gambetta}, \citenamefont {Houck},
  \citenamefont {Schuster}, \citenamefont {Majer}, \citenamefont {Blais},
  \citenamefont {Devoret}, \citenamefont {Girvin},\ and\ \citenamefont
  {Schoelkopf}}]{Koch2007}%
  \BibitemOpen
  \bibfield  {author} {\bibinfo {author} {\bibfnamefont {Jens}\ \bibnamefont
  {Koch}}, \bibinfo {author} {\bibfnamefont {Terri~M.}\ \bibnamefont {Yu}},
  \bibinfo {author} {\bibfnamefont {Jay}\ \bibnamefont {Gambetta}}, \bibinfo
  {author} {\bibfnamefont {A.~A.}\ \bibnamefont {Houck}}, \bibinfo {author}
  {\bibfnamefont {D.~I.}\ \bibnamefont {Schuster}}, \bibinfo {author}
  {\bibfnamefont {J.}~\bibnamefont {Majer}}, \bibinfo {author} {\bibfnamefont
  {Alexandre}\ \bibnamefont {Blais}}, \bibinfo {author} {\bibfnamefont {M.~H.}\
  \bibnamefont {Devoret}}, \bibinfo {author} {\bibfnamefont {S.~M.}\
  \bibnamefont {Girvin}}, \ and\ \bibinfo {author} {\bibfnamefont {R.~J.}\
  \bibnamefont {Schoelkopf}},\ }\bibfield  {title} {\enquote {\bibinfo {title}
  {Charge-insensitive qubit design derived from the cooper pair box},}\ }\href
  {\doibase 10.1103/PhysRevA.76.042319} {\bibfield  {journal} {\bibinfo
  {journal} {Phys. Rev. A}\ }\textbf {\bibinfo {volume} {76}},\ \bibinfo
  {pages} {042319} (\bibinfo {year} {2007})}\BibitemShut {NoStop}%
\bibitem [{\citenamefont {Wu}\ \emph {et~al.}(2017)\citenamefont {Wu},
  \citenamefont {Long}, \citenamefont {Ku}, \citenamefont {Lake}, \citenamefont
  {Bal},\ and\ \citenamefont {Pappas}}]{wu2017overlap}%
  \BibitemOpen
  \bibfield  {author} {\bibinfo {author} {\bibfnamefont {X}~\bibnamefont {Wu}},
  \bibinfo {author} {\bibfnamefont {JL}~\bibnamefont {Long}}, \bibinfo {author}
  {\bibfnamefont {HS}~\bibnamefont {Ku}}, \bibinfo {author} {\bibfnamefont
  {RE}~\bibnamefont {Lake}}, \bibinfo {author} {\bibfnamefont {M}~\bibnamefont
  {Bal}}, \ and\ \bibinfo {author} {\bibfnamefont {DP}~\bibnamefont {Pappas}},\
  }\bibfield  {title} {\enquote {\bibinfo {title} {Overlap junctions for high
  coherence superconducting qubits},}\ }\href@noop {} {\bibfield  {journal}
  {\bibinfo  {journal} {Applied Physics Letters}\ }\textbf {\bibinfo {volume}
  {111}},\ \bibinfo {pages} {032602} (\bibinfo {year} {2017})}\BibitemShut
  {NoStop}%
\bibitem [{\citenamefont {Long}(2020)}]{Junling2020Superconducting}%
  \BibitemOpen
  \bibfield  {author} {\bibinfo {author} {\bibfnamefont {Junling}\ \bibnamefont
  {Long}},\ }\href@noop {} {\emph {\bibinfo {title} {Superconducting Quantum
  Circuits for QuantumInformation Processing}}}\ (\bibinfo  {publisher}
  {University of Colorado Boulder},\ \bibinfo {year} {2020})\BibitemShut
  {NoStop}%
\bibitem [{\citenamefont {Barnes}\ and\ \citenamefont
  {Sarma}(2012)}]{barnes2012analytically}%
  \BibitemOpen
  \bibfield  {author} {\bibinfo {author} {\bibfnamefont {Edwin}\ \bibnamefont
  {Barnes}}\ and\ \bibinfo {author} {\bibfnamefont {S~Das}\ \bibnamefont
  {Sarma}},\ }\bibfield  {title} {\enquote {\bibinfo {title} {Analytically
  solvable driven time-dependent two-level quantum systems},}\ }\href@noop {}
  {\bibfield  {journal} {\bibinfo  {journal} {Physical review letters}\
  }\textbf {\bibinfo {volume} {109}},\ \bibinfo {pages} {060401} (\bibinfo
  {year} {2012})}\BibitemShut {NoStop}%
\bibitem [{\citenamefont {Premaratne}\ \emph {et~al.}(2019)\citenamefont
  {Premaratne}, \citenamefont {Yeh}, \citenamefont {Wellstood},\ and\
  \citenamefont {Palmer}}]{Premaratne2019}%
  \BibitemOpen
  \bibfield  {author} {\bibinfo {author} {\bibfnamefont {Shavindra~P.}\
  \bibnamefont {Premaratne}}, \bibinfo {author} {\bibfnamefont {Jen-Hao}\
  \bibnamefont {Yeh}}, \bibinfo {author} {\bibfnamefont {F.~C.}\ \bibnamefont
  {Wellstood}}, \ and\ \bibinfo {author} {\bibfnamefont {B.~S.}\ \bibnamefont
  {Palmer}},\ }\bibfield  {title} {\enquote {\bibinfo {title} {Implementation
  of a generalized controlled-not gate between fixed-frequency transmons},}\
  }\href {\doibase 10.1103/PhysRevA.99.012317} {\bibfield  {journal} {\bibinfo
  {journal} {Phys. Rev. A}\ }\textbf {\bibinfo {volume} {99}},\ \bibinfo
  {pages} {012317} (\bibinfo {year} {2019})}\BibitemShut {NoStop}%
\bibitem [{\citenamefont {Knill}\ \emph {et~al.}(2008)\citenamefont {Knill},
  \citenamefont {Leibfried}, \citenamefont {Reichle}, \citenamefont {Britton},
  \citenamefont {Blakestad}, \citenamefont {Jost}, \citenamefont {Langer},
  \citenamefont {Ozeri}, \citenamefont {Seidelin},\ and\ \citenamefont
  {Wineland}}]{knill2008randomized}%
  \BibitemOpen
  \bibfield  {author} {\bibinfo {author} {\bibfnamefont {Emanuel}\ \bibnamefont
  {Knill}}, \bibinfo {author} {\bibfnamefont {Dietrich}\ \bibnamefont
  {Leibfried}}, \bibinfo {author} {\bibfnamefont {Rolf}\ \bibnamefont
  {Reichle}}, \bibinfo {author} {\bibfnamefont {Joe}\ \bibnamefont {Britton}},
  \bibinfo {author} {\bibfnamefont {R~Brad}\ \bibnamefont {Blakestad}},
  \bibinfo {author} {\bibfnamefont {John~D}\ \bibnamefont {Jost}}, \bibinfo
  {author} {\bibfnamefont {Chris}\ \bibnamefont {Langer}}, \bibinfo {author}
  {\bibfnamefont {Roee}\ \bibnamefont {Ozeri}}, \bibinfo {author}
  {\bibfnamefont {Signe}\ \bibnamefont {Seidelin}}, \ and\ \bibinfo {author}
  {\bibfnamefont {David~J}\ \bibnamefont {Wineland}},\ }\bibfield  {title}
  {\enquote {\bibinfo {title} {Randomized benchmarking of quantum gates},}\
  }\href@noop {} {\bibfield  {journal} {\bibinfo  {journal} {Physical Review
  A}\ }\textbf {\bibinfo {volume} {77}},\ \bibinfo {pages} {012307} (\bibinfo
  {year} {2008})}\BibitemShut {NoStop}%
\bibitem [{\citenamefont {O'Brien}\ \emph {et~al.}(2004)\citenamefont
  {O'Brien}, \citenamefont {Pryde}, \citenamefont {Gilchrist}, \citenamefont
  {James}, \citenamefont {Langford}, \citenamefont {Ralph},\ and\ \citenamefont
  {White}}]{o2004quantum}%
  \BibitemOpen
  \bibfield  {author} {\bibinfo {author} {\bibfnamefont {Jeremy~L}\
  \bibnamefont {O'Brien}}, \bibinfo {author} {\bibfnamefont {GJ}~\bibnamefont
  {Pryde}}, \bibinfo {author} {\bibfnamefont {Alexei}\ \bibnamefont
  {Gilchrist}}, \bibinfo {author} {\bibfnamefont {DFV}\ \bibnamefont {James}},
  \bibinfo {author} {\bibfnamefont {Nathan~K}\ \bibnamefont {Langford}},
  \bibinfo {author} {\bibfnamefont {TC}~\bibnamefont {Ralph}}, \ and\ \bibinfo
  {author} {\bibfnamefont {AG}~\bibnamefont {White}},\ }\bibfield  {title}
  {\enquote {\bibinfo {title} {Quantum process tomography of a controlled-not
  gate},}\ }\href@noop {} {\bibfield  {journal} {\bibinfo  {journal} {Physical
  review letters}\ }\textbf {\bibinfo {volume} {93}},\ \bibinfo {pages}
  {080502} (\bibinfo {year} {2004})}\BibitemShut {NoStop}%
\bibitem [{\citenamefont {Dewes}\ \emph {et~al.}(2012)\citenamefont {Dewes},
  \citenamefont {Ong}, \citenamefont {Schmitt}, \citenamefont {Lauro},
  \citenamefont {Boulant}, \citenamefont {Bertet}, \citenamefont {Vion},\ and\
  \citenamefont {Esteve}}]{dewes2012characterization}%
  \BibitemOpen
  \bibfield  {author} {\bibinfo {author} {\bibfnamefont {A}~\bibnamefont
  {Dewes}}, \bibinfo {author} {\bibfnamefont {FR}~\bibnamefont {Ong}}, \bibinfo
  {author} {\bibfnamefont {V}~\bibnamefont {Schmitt}}, \bibinfo {author}
  {\bibfnamefont {R}~\bibnamefont {Lauro}}, \bibinfo {author} {\bibfnamefont
  {N}~\bibnamefont {Boulant}}, \bibinfo {author} {\bibfnamefont
  {P}~\bibnamefont {Bertet}}, \bibinfo {author} {\bibfnamefont {D}~\bibnamefont
  {Vion}}, \ and\ \bibinfo {author} {\bibfnamefont {D}~\bibnamefont {Esteve}},\
  }\bibfield  {title} {\enquote {\bibinfo {title} {Characterization of a
  two-transmon processor with individual single-shot qubit readout},}\
  }\href@noop {} {\bibfield  {journal} {\bibinfo  {journal} {Physical review
  letters}\ }\textbf {\bibinfo {volume} {108}},\ \bibinfo {pages} {057002}
  (\bibinfo {year} {2012})}\BibitemShut {NoStop}%
\bibitem [{\citenamefont {Kaye}\ \emph {et~al.}(2007)\citenamefont {Kaye},
  \citenamefont {Laflamme}, \citenamefont {Mosca} \emph
  {et~al.}}]{kaye2007introduction}%
  \BibitemOpen
  \bibfield  {author} {\bibinfo {author} {\bibfnamefont {Phillip}\ \bibnamefont
  {Kaye}}, \bibinfo {author} {\bibfnamefont {Raymond}\ \bibnamefont
  {Laflamme}}, \bibinfo {author} {\bibfnamefont {Michele}\ \bibnamefont
  {Mosca}},  \emph {et~al.},\ }\href@noop {} {\emph {\bibinfo {title} {An
  introduction to quantum computing}}}\ (\bibinfo  {publisher} {Oxford
  university press},\ \bibinfo {year} {2007})\BibitemShut {NoStop}%
\end{thebibliography}%


%

\end{document}